\documentclass[english,twocolumn,superscriptaddress,aps,nofootinbib]{revtex4-1}
\usepackage{amsmath}
\usepackage[english]{babel}
\usepackage{graphicx}
\usepackage{epstopdf}
\usepackage{amssymb}
\usepackage{amsfonts}
\usepackage[mathcal]{eucal}
\usepackage[colorlinks,citecolor=blue,linkcolor=blue,urlcolor=blue]{hyperref}
\usepackage{txfonts}
\usepackage[tiny,center,uppercase]{titlesec}

\begin{document}

\newcommand{\bt}{\textbf}
\newcommand{\ep}{\varepsilon}
\newcommand{\vp}{\varphi}
\newcommand{\bs}{\boldsymbol}
\newcommand{\te}{\theta}
\newcommand{\vk}{\varkappa}
\newcommand{\si}{\sigma}
\newcommand{\om}{\omega}
\newcommand{\cd}{\!\cdot\!}
\newcommand{\op}[1]{\boldsymbol{\mathfrak{#1}}}
\newcommand{\fp}[2]{(#1\!\cdot\! #2)}
\newcommand{\thp}[2]{\bs #1\!\cdot\!\bs #2}
\newcommand{\opA}[1]{\boldsymbol{\mathsf{#1}}}
\newcommand*{\ch}{\mathop{}\!\mathrm{ch}}
\newcommand*{\sh}{\mathop{}\!\mathrm{sh}}
\newcommand*{\tah}{\mathop{}\!\mathrm{th}}
\setlength{\abovedisplayskip}{1.5pt}
\setlength{\belowdisplayskip}{1.5pt}
\titlespacing{\section}{0pt}{10pt}{5pt}
\title{Collapse-and-revival dynamics of strongly laser-driven electrons}

\author{O.\ D.\ Skoromnik}
\email[Corresponding author: ]{olegskor@gmail.com}
\affiliation{Max Planck Institute for Nuclear Physics, Saupfercheckweg 1, 69117 Heidelberg, Germany}
\author{I.\ D.\ Feranchuk}
\affiliation{Belarusian State University, 4 Nezavisimosty Ave., 220030, Minsk,   Belarus}
\author{C. H. Keitel}
\affiliation{Max Planck Institute for Nuclear Physics, Saupfercheckweg 1, 69117 Heidelberg, Germany}

\begin{abstract}
The relativistic quantum dynamics of an electron in an intense single-mode quantized electromagnetic field is investigated with special emphasis on the spin degree of freedom. In addition to fast spin oscillations at the laser frequency, a second time scale is identified due to the intensity dependent emissions and absorptions of field quanta. In analogy to the well-known phenomenon in atoms at moderate laser intensity, we put forward the conditions of collapses and revivals for the spin evolution in laser-driven electrons starting at feasible $10^{18}$ W/cm$^2$.
\end{abstract}

\maketitle

PACS number(s): 12.20.-m, 13.88.+e

\section{Introduction}\label{intro}
The exact solution of the Dirac equation for an electron in a classical field of a plane electromagnetic wave was found by Volkov \cite{Volkov}. This solution initiated numerous investigations of quantum electrodynamical (QED) processes in the presence of a strong electromagnetic field. The fundamental basis for this part of QED was developed in a series of papers \cite{Nikishov1,*Goldman,*Ritus,Reiss,Ritus1985} about fifty years ago. Nevertheless, with the development of powerful lasers providing intensities up to $I\sim10^{22}\  \textrm{W}/\textrm{cm}^2$  \cite{PhelixFacility,*Gerstner,*Yanovsky} and ambitions plans beyond \cite{ELI,*HIPER}, the analysis of various quantum phenomena \cite{KeitelReview,*Ehlotzky,*Mourou,*Marklund.RevModPhys.78.591,*Salamin200641} such as multi-photon Compton scattering \cite{Karlovets,Piazza,*BocaOprea,*Seipt,*BocaFlorescu,*HarveyHeinzl,Gavrila,*BergouVarro}, electron--positron pair creation \cite{Titov,*Mueller,*DiPiazza-Milstein}, spin effects \cite{PhysRevA.87.023418,*Karlovets.PhysRevA.86.062102,*Ahrens.PhysRevLett.109.043601,*Meuren,*Faisal,*Walser,*Keitel.PhysRevLett.83.4709}, and quantum plasmas \cite{Shukla,*Marklund} has recently become increasingly relevant. In all this research, however, the external electromagnetic field is considered as classical, and its quantum fluctuations are generally believed to be negligible for interactions with very strong fields. The scattering amplitude for these types of processes in Furry's representation is defined by the same Feynman  diagrams as in vacuum QED, however the exact solutions of the Dirac equation in the presence of a plane electromagnetic wave are used for the external electron lines \cite{LandauQED}.

At the same time, it is well known from quantum optics that the dynamics of an atom in a comparably weaker and resonant laser field depends on the quantum fluctuations of the field. One of the most interesting phenomena of this kind is the collapse--revival effect in the evolution of the Jaynes--Cummings model \cite{JaynesCummings,*Cummings,Knight} for a two-level atom. This effect was predicted theoretically \cite{Eberly1,*Eberly2,*Rempe1,*Schleich} and later observed experimentally \cite{Rempe2,*Brune}. Its qualitative explanation and analytical description was also given by \cite{Leonov,*Feranchuk2011,*Feranchuk2009,Scully,Filipowicz2}. It was shown that the evolution of the population of the atomic states is characterized by two time scales. The first time scale is the period of the Rabi oscillations while the second  slower one is defined by the collapse and revival times of the populations being associated with the absorption and emission of the field quanta.

An electron can on one hand be considered as a two-level system with regard to spin space. On the other hand the electron has no other internal quantum degrees of freedom such as the atom and especially strong laser-electron dynamics is generally considered to be almost perfectly described within a classical picture \cite{Keitel.PhysRevA.72.033402}. For that reason, the questions of the feasibility of collapse--revival dynamics for laser-driven electrons arises. In this paper, we present the electron spin evolution in a single-mode quantized field within the framework of Dirac theory. We show that an electron may exhibit collapse and revival dynamics similar to that of two-level atoms and demonstrate a possible implementation of this pure quantum effect already at widely available laser intensities in the moderately relativistic regime. 

Qualitatively different time scales can be selected in the evolution of a quasi-energy electron state $\psi^{(e)}_p (\bs r, t)$ \cite{Zeldovich} with definite quasi-momentum $p$. One of these scales is defined by the frequency $\om$ of the electromagnetic field, i. e. in natural units $\hbar = c = 1$ as used in the complete article:
\begin{eqnarray*}
	T = \frac{2 \pi}{\omega} \approx 10^{-4} \ \textrm{cm},
\end{eqnarray*}
for the photon energy  $\om \approx 1\, \textrm{eV}$.

Another scale is defined by the coherence time $T_c$ which was introduced in \cite{Ritus, Ritus1985}. The characteristic time $T_c$ is inversely proportional to the probability $\mathsf{w}_c$ of a photon emission per unit of time by an electron which is in a Volkov state.  It equals the  distance in which the uncertainty of the phase for the Volkov wave function changes by   $2\pi$.   An approximation for this time was found in \cite{Ritus1985}:
\begin{eqnarray*}
T_c \sim \mathsf{w}_c^{-1} \approx \frac{2}{\xi^2}T; \quad \xi = \frac{e a}{m} \leq 1,	
\end{eqnarray*}
where $\xi$ is an invariant parameter which characterizes the ``strength'' of the electromagnetic field \cite{LandauQED}, $a$ is the amplitude of the electromagnetic field potential, and $e$ and $m$ are the electronic charge and mass, respectively. It is well known that the value $\xi \approx 0.35$   corresponds to the intensity of electromagnetic field $I_0\approx 10^{18}\, \textrm{W}/\textrm{cm}^2$ when $\om \approx 1\, \textrm{eV}$.  

The existence of two time scales is expected to be observable   when the  travel time of an electron in the field $L=T_{\mathrm{int}}$ (that is, the time of interaction between an electron and the field) satisfies
\begin{eqnarray*}
	L > T_c \gg T,
\end{eqnarray*}
which can be fulfilled for realizable parameters of the laser pulse.

In a more general view, the slow oscillations are characterized by the radiation shift of the electron's quasi-energy, which is defined via the Dirac equation with mass operator \cite{Akhiezer}. The shift appears both due to the quantum nature of the driving electromagnetic field and the environmental vacuum field. However, the laser-driven processes of emission and absorption of photons with frequency $\om$ are the main contribution to the transition amplitudes that are proportional to the number of quanta in the radiation mode \cite{LandauQED}. Therefore, the system of an electron and a one mode quantized field can be investigated as a closed system in a non-perturbative way.

The present paper is organized as follows. In Sec. II, the solution of the Dirac equation for a one-mode quantized field is found directly in the operator form because it is more suitable for the problem under consideration. In this solution, the state vector of the system is expressed in terms of the field operators without Bargman's representation. In Sec. III, the evolution of the spin operator is found and all necessary matrix elements are calculated with the assumption that at the initial moment of time (when the electron enters the laser pulse) the state vector of the system consists of a free electron wave function and a coherent state for the driving electromagnetic field. It is shown that the evolution of the spin is governed by two scales employing parameters which are in agreement with all used approximations. In Secs. IV and V the possibility of an experimental observation of the effect is discussed and an explicit proof of the quasi-classical limit of the quantum case is presented. In Appendix A the validity of the single-mode approximation is investigated and justified for the relevant range of parameters. In Appendixes B, C, D and E technical details of the calculations are presented.

\section{Solution of the Dirac equation in a quantized electromagnetic field}\label{sec2}

The accurate analytic solution of the Dirac equation in a single-mode quantized electromagnetic field  was found by Berson \cite{Berson, *[{see also }] Filipowicz1,*Filipowicz3} in 1969. However, that solution was obtained in a coordinate (Bargmann) representation for the creation and annihilation operators of the field. In what follows we show that it is possible to find an analogous solution directly in operator form. For that reason we start with a QED Schr\"{o}dinger equation with a single-mode quantized field (the justification of the single-mode approximation for a multi-mode laser pulse is referred to Appendix A)
\begin{equation}\label{1}
i\frac{\partial \Psi}{\partial t} = \left(\om a^\dag a + \bs \alpha\cdot(\bs p - e\bs A) + \beta m + e\phi\right)\Psi,
\end{equation}
with the potential
\begin{equation}\label{2}
\bs A = \frac{\bs e}{\sqrt{2\omega V}}\left(a e^{i\thp{k}{r}}+a^\dag e^{-i\thp{k}{r}}\right),
\end{equation}
Dirac matrices $\bs \alpha$ and $\beta$, a polarization vector $\bs e$, a normalization volume $V$, photon four-vector $k= (k^0 = \om, \bs k) $, photon annihilation and creation operators $a$ and $a^\dag$ of the laser mode, an electron charge $e$ and mass $m$. The equation (\ref{1}) can be written in the covariant form of the Dirac equation if the transformation $\Psi = e^{-i \omega t a^\dag a}\psi$ is used. 

As a consequence the covariant form reads
\begin{eqnarray}\label{3}
	\left(i\gamma^\mu \partial_\mu - \gamma^\mu eA_\mu - m\right)\psi = 0,
\end{eqnarray}
with the four product defined as $\fp{k}{x} = k^0 t - \thp{k}{x}$, $\partial_\mu = \partial/\partial x^\mu $, the metric tensor $g^{\mu\nu} = \mathrm{diag}(1,-1,-1,-1)$ and the four-potential of the field
\begin{eqnarray*}
	A_\mu = \frac{e_\mu}{\sqrt{2\omega V}}\left(a e^{-i\fp{k}{r}}+a^\dag e^{i\fp{k}{r}}\right).
\end{eqnarray*}

Now we perform the transformation $\psi = e^{i\fp{k}{x}a^\dag a}\chi$ in order to exclude the electron coordinates from the field operators. As a result, the operators are transformed as follows:
\begin{eqnarray}\label{4}
i\gamma^\mu\partial_\mu  \rightarrow i\gamma^\mu\partial_\mu - \hat k a^\dag a, \nonumber
\\
a \rightarrow ae^{i\fp{k}{x}},\quad a^\dag \rightarrow a^\dag e^{-i\fp{k}{x}},
\end{eqnarray}
\noindent and one can find for Eq. (\ref{1})
\begin{equation}\label{5}
\left(i\hat\partial - \hat k a^\dag a - \hat b (a+a^\dag) - m\right)\chi = 0,
\end{equation}
\noindent where $b_\mu = e e_\mu/\sqrt{2V\om}$ and    $\hat f\equiv \gamma^\mu f_\mu$ for any four-vector $f$.

In the general case, the solution of Eq. (\ref{5}) has the form
\begin{equation}\label{6}
\chi = e^{-i\fp{q}{x}}\phi,
\end{equation}
where $q^\mu$ is the constant four-vector and   the  state vector $\phi$   satisfies
\begin{equation}\label{7}
\hat H \phi \equiv \left(\hat q -\hat k a^\dag a- \hat b (a+a^\dag)-m\right)\phi = 0.
\end{equation}

To  solve Eq. (\ref{7}), the photon and spin variable should be separated, which can be performed by means of the transformation
\begin{eqnarray*}
	\phi = U\vp, \quad U = e^{l\hat k\hat b (a+a^\dag)},
\end{eqnarray*}
with a constant $l$ that is to be defined later on.

In the Lorentz gauge, the value $\fp{k}{b} = 0$ leads to
\begin{eqnarray*}
	\hat b \hat k + \hat k \hat b = 2\fp{k}{b} = 0, \quad \hat k\hat k=k^2=0,
	\\
	U = e^{l\hat k\hat b (a+a^\dag)}= 1+l \hat k \hat b (a+a^\dag).
\end{eqnarray*}

Calculating the operator $\hat H' = U^{-1}  \hat H U$,   one can find
\begin{align}\label{8}
&\Big(\hat q - \hat k a^\dag a +  l(a+a^\dag)(2\hat b \fp{q}{k}-2\hat k \fp{q}{b}) - \nonumber
\\ 
&- \hat b (a+a^\dag) - 2l^2(a+a^\dag)^2\fp{q}{k}b^2 \hat k - m + l (a+a^\dag)^22 \fp{q}{k}b^2 \hat k\Big)\varphi \nonumber
\\
&= 0.
\end{align}
If we choose
\begin{eqnarray*}
	l = 1/(2\fp{q}{k}),
\end{eqnarray*}
the terms linear in $\hat b$ are canceled  and Eq. (\ref{8}) changes to
\begin{eqnarray}\label{9}
\hat H'\vp =  \Bigg(\hat q - m -  \hat k  [a^\dag a + \nonumber\\+    \frac{\fp{q}{b}}{\fp{q}{k}}(a+a^\dag) - \frac{b^2}{2\fp{q}{k}}(a+a^\dag)^2 ]
\Bigg)\vp = 0.
\end{eqnarray}
The operator $\hat H'$ is diagonalized via the following transformations
\begin{eqnarray}\label{10}
\varphi = D B \Theta,\quad D = e^{\alpha a^\dag - \alpha^* a}, \quad  B = e^{-\frac{\eta}{2}(a^2-a^{\dag2})},
\end{eqnarray}
with the parameters $\alpha$ and $\eta$.
Here, as well known, the operator $D$ shifts $a$ and $a^\dag$ by the complex numbers $\alpha$ and $\alpha^*$, respectively:
\begin{eqnarray*}
	D^{-1}a D = a +\alpha, \quad D^{-1}a^\dag D= a^\dag + \alpha^*.
\end{eqnarray*}
The operator $B$ transforms the operators $a$ and $a^\dag$ as follows
\begin{eqnarray*}
B^{-1}a B =  a \ch \eta + a^\dag \sh \eta, 
\\
B^{-1} a^\dag B =  a \sh \eta + a^\dag \ch \eta. 
\end{eqnarray*}

These parameters are defined by the condition that the operator $\hat H_1 =B^{-1} D^{-1}\hat H' D B $ transforms to a diagonal form. This leads to
\begin{eqnarray}\label{11}
\hat H_1 \Theta &=&  \Bigg(\hat q - \hat k \Bigg(\sqrt{1-\frac{2b^2}{\fp{q}{k}}}(a^\dag a +\frac{1}{2})-\frac{1}{2} \nonumber
\\
&-& \frac{\fp{q}{b}^2}{\fp{q}{k} }\frac{1}{\fp{q}{k} - 2b^2 }\Bigg) - m\Bigg)\Theta = 0; \nonumber
\\
\alpha &=& - \frac{\fp{q}{b}}{\fp{q}{k}}\frac{1}{1 - {2b^2}/\fp{q}{k}}, \nonumber
\\
\ch\eta&=&\frac{1}{2}\left(\sqrt{\vk}+\frac{1}{\sqrt{\vk}}\right),\quad \vk = \frac{1}{\sqrt{1-\frac{2b^2}{\fp{q}{k}}}}.
\end{eqnarray}

The eigenvector of Eq. (\ref{11}) can be represented in the form
\begin{equation}\label{12}
(\hat p_n - m)\Theta = 0, \quad \Theta  = u(p_n)|n\rangle,
\end{equation}
where $|n\rangle$ is the state vector of the harmonic oscillator, $u(p_n)$ is the constant bispinor which satisfies the same equation as in the case of the free electron, and the  vector $p_n$   depends on the quantum number $n$ as follows
\begin{eqnarray}\label{13}
p_n = q-k \Bigg(\sqrt{1-\frac{2b^2}{\fp{q}{k}}}(n +\frac{1}{2})-\frac{1}{2} - \nonumber
\\
- \frac{\fp{q}{b}^2}{\fp{q}{k}^2}\frac{1}{1 - 2b^2/\fp{q}{k}}\Bigg).
\end{eqnarray}
As a result of all these transformations, the wave function of the electron in the single-mode quantized field has the following form
\begin{eqnarray}\label{14}
\psi_{qn} = N e^{-i\fp{q}{x}+ia^\dag a \fp{k}{x}}\left(1+\frac{\hat k \hat b}{2\fp{q}{k}}(a+a^\dag)\right)\nonumber
\\
\cdot e^{\alpha (a^\dag -a)} e^{-\frac{\eta}{2}(a^2-a^{\dag 2})}u(p_n)|n\rangle,
\end{eqnarray}
where $N$ is a normalization constant.

The vector $p_n$ satisfies
\begin{equation}\label{15}
p_n^2 - m^2 = 0,
\end{equation}
which is a consequence of Eq. (\ref{12}), and  the four-vector $q$ is the total moment of the system.

The wave function (\ref{14}) coincides with Berson's solution \cite{Berson} if the Bargmann representation
\begin{equation}\label{16}
a = \frac{1}{\sqrt{2}}(x + \frac{\partial}{\partial x}); \ a^\dag = \frac{1}{\sqrt{2}}(x - \frac{\partial}{\partial x}),
\end{equation}
is used for the operators, with $x$ being the field variable.

It was also shown in \cite{Berson} that if the field operators are changed to the classical values $ a \approx a^\dag \approx \beta $, the wave function (\ref{14}) coincides with Volkov's solution \cite{Volkov} for a classical field,
\begin{equation}\label{17}
A_{\mu} = \frac{e_\mu}{\sqrt{2\omega V}}2 \beta \cos \fp{k}{x}.
\end{equation}
The wave functions $\psi_{qn}$ satisfy the orthogonality condition
\begin{equation}\label{18}
\frac{1}{(2\pi)^3 }\int \psi^\dag_{q^\prime n^\prime}\psi_{qn}d\bs r =N^2 2\ep_n \delta_{n^\prime n}\delta (\bs q^\prime-\bs q),
\end{equation}
where $\ep_n = \sqrt{p_n^2+m^2}$. Thus the normalization constant $N$ can be chosen   as for the free electron   $N=1/\sqrt{2\ep_n}$.

In order to find the evolution of the system state vector,  one should fix the time reference. It is natural to connect it with the moment $t = 0$ when the electron passes the boundary of the laser pulse. This means that the system state vector at $t = 0$ is described by a free electron wave function and the field by a coherent state \cite{Scully}
\begin{eqnarray}\label{19}
\psi_0 = e^{i\thp{p_0}{(\bs r - \bs r_0)}}\frac{u(p_0)}{\sqrt{2\ep_0}}|\beta\rangle, \quad |\beta\rangle = \sum_{n=0}^{\infty} \frac{\beta^n}{\sqrt{n!}}| n \rangle e^{- \beta^2/2},
\end{eqnarray}
where $|\beta\rangle$ is a coherent state of the field, $\bs p_0$ the electron momentum, $u(p_0)$ a constant bispinor normalized with the condition
\begin{eqnarray*}
	\bar u(p_0)\gamma^0 u(p_0) = 2\ep_0,
\end{eqnarray*}
and the vector $\bs r_0$ defines the initial phase of the electron state.

Let us now use a linear combination of the solutions (\ref{14})
\begin{eqnarray}\label{20}
\Psi (x) = \int d\bs q \sum_n C_{\bs q,n}\psi_{qn}(x); \ x = (t, \bs x)
\end{eqnarray}
in order to satisfy the initial condition
\begin{eqnarray}\label{21}
e^{i\bs{p_0}(\bs r - \bs r_0)}\frac{u(p_0)} {\sqrt{2\ep_0}}|\beta\rangle = \Psi\big|_{t=0}.
\end{eqnarray}
The wave functions $\psi_{qn}$ are orthogonal, such that Eq. (\ref{21}) leads to the following expression for the coefficient	$C_{\bs q, n}$
\begin{eqnarray}\label{22}
C_{\bs q,n} = \frac{\bar u(p_n)}{\sqrt{2\ep_n}}\gamma^0\frac{u(p_0)}{\sqrt{2\ep_0}}\left(1+\frac{2\alpha}{\fp{q}{k}}\right)M_{\bs q,n}+\frac{\bar u(p_n)}{\sqrt{2\ep_n}}\hat b\hat k\gamma^0 \nonumber
\\
\times
\frac{u(p_0)}{\sqrt{2\ep_0}}\frac{\sqrt{\vk}}{2\fp{q}{k}}\Big(\sqrt{n+1}M_{\bs q,n+1}+\sqrt{n}M_{\bs q,n-1}\Big),
\end{eqnarray}
where the matrix element $M_{\bs q,n}$ is calculated via
\begin{align}\label{23}
M_{\bs q,n} =\ & \frac{1}{(2\pi)^3}\int d\bs r e^{-i(\bs q-\bs p_0)\cdot( \bs r - \bs r_0 )} \nonumber
\\
&\cdot\frac{e^{-|\theta|^2/2 + \alpha \theta -\alpha^2/2+1/2(\theta -\alpha)^2 \textrm{th}\eta}}{\sqrt{\textrm{ch}\eta}}  \frac{1}{\sqrt{n!}}\left(\frac{\textrm{th}\eta}{2}\right)^{\frac{n}{2}} \nonumber
\\
&\cdot H_n\left(\frac{\theta-\alpha}{\sqrt{2\textrm{th}\eta}\textrm{ch}\eta}\right),
\end{align}
$\theta = \beta e^{i\thp{k}{r}}$, and $\beta$ is a coherent state parameter. The details of the calculation of the coefficients can be found in Appendix B.

For that reason, the wave function (\ref{20}) with the coefficients (\ref{22}) describes exactly the evolution of the system consisting of a relativistic electron in  a linearly polarized single-mode quantized field.
Our purpose is to analyze the influence of quantum effects on the  dynamics of the observable values for this  system, and it is important to estimate the characteristic parameters of the problem. Consider a strong laser field, with density of photons \cite{Ritus}
\begin{eqnarray*}
	\rho = \frac{n}{V},
\end{eqnarray*}
with $n$ being the number of photons and $V$ a normalization volume. For real system parameters, the limits
\begin{eqnarray}\label{24}
V \rightarrow \infty; \ n \rightarrow \infty; \ n/V \rightarrow \mathrm{const}.
\end{eqnarray}
should be considered and all other terms inversely proportional to a power of $V$ can be neglected.

In spite  of the fact that the photon energy is small compared to the electron energy $\om/\ep \ll 1$,   the total momentum of the field $\bs k n$ can be  compared with the momentum of the electron $\bs p_0 \sim \bs k n $, because the photon number $n$ is large.

Let us now estimate the parameters $\alpha$, $\ch\eta$, $\sh\eta$, $\tah\eta$ in the state vector (\ref{14}) taking into account of the condition (\ref{24}):
\begin{align}\label{25}
&\alpha = -\frac{\fp{q}{b}}{\fp{q}{k}}\frac{1}{1-\frac{2b^2}{\fp{q}{k}}}\approx -\frac{\fp{q}{b}}{\fp{q}{k}},\nonumber
\\
&\vk =  \frac{1}{\sqrt{1-\frac{2b^2}{\fp{q}{k}}}} \approx 1, \nonumber
\\
\ch &\eta = \frac{1}{2}\sqrt[4]{1-\frac{2b^2}{\fp{q}{k}}}\left(\frac{\sqrt{1-\frac{2b^2}{\fp{q}{k}}}+1}{\sqrt{1-\frac{2b^2}{\fp{q}{k}}}}\right)\approx 1,\nonumber
\\
\tah &\eta = \frac{1-\sqrt{1-\frac{2b^2}{\fp{q}{k}}}}{1+\sqrt{1-\frac{2b^2}{\fp{q}{k}}}}\approx \sh \eta \approx \eta \approx \frac{b^2}{2\fp{q}{k}}.
\end{align}

It is also important   to find the dispersion relation for the zero  component $q^0$  of the four-vector $q$ that is given by the equation
\begin{eqnarray}\label{26}
p_n^2 - m^2 = 0.
\end{eqnarray}
Substituting into Eq. (\ref{26}) the connection between the vector $p_n$ and $q$, the dispersion relation can be found:
\begin{eqnarray}\label{27}
	&(q^0)^2 - 2 q^0 \tilde k^0 - (\bs q^2 - 2\bs q \cdot  \tilde{\bs k} + m^2) = 0,
	\\
	&\tilde k = k^0 \left(\sqrt{1-\frac{2b^2}{\fp{q}{k}}}\left(n+\frac{1}{2}\right)-\frac{1}{2}-\frac{\fp{q}{b}^2}{\fp{q}{k}^2}\frac{1}{1-2b^2/\fp{q}{k}^2}\right). \nonumber
\end{eqnarray}
Then the solution of the quadratic equation gives the required zero component of the four-vector $q$ in the limits (\ref{24})
\begin{equation}\label{28}
q^0 = \omega n+\sqrt{m^2+(\bs q-\bs kn)^2}.
\end{equation}


\section{Electron spin dynamics in a single-mode quantized field}\label{sec3}
In order to analyze the influence of quantum effects on the system dynamics, we consider  the electron spin four-vector, which is defined via \cite{LandauQED}
\begin{equation}\label{29}
s^\mu(\bs x ,t) = \frac{\langle\psi|\gamma^0\gamma^5\gamma^\mu\delta(\bs x  - \bs r^\prime)|\psi\rangle}{\langle\psi|\psi\rangle}.
\end{equation}

To calculate the average value of the spin (\ref{29}), one should perform the averaging in spin space. For this purpose, the density matrix of the free electron is used,
\begin{eqnarray}\label{30}
\rho =u(p)\otimes\bar u(p)= \frac{1}{2}(\hat p + m)(1-\gamma^5\hat a),
\end{eqnarray}
where $a$ is the four-vector that differs from the four-vector $s$ by the normalization $a = \frac{\ep}{m} s$ \cite{LandauQED}.
In the case of a free electron, the four-vector $s^\mu$ has components
\begin{eqnarray}\label{31}
\bs s = \frac{m}{\ep}\bs \zeta +\frac{\bs p(\thp{p}{\zeta})}{\ep(\ep+m)}, \quad s^0 = \frac{\thp{p}{\zeta}}{\ep},
\end{eqnarray}
where $\bs \zeta$ is the electron spin in the rest frame
\begin{eqnarray*}
	\bs \zeta = \langle\bs\si\rangle,
\end{eqnarray*}
$\bs\si$ are the Pauli matrices, $\bs p$ is the electron momentum, and $\ep$ the electron energy, which satisfies
\begin{eqnarray*}
	\ep^2 = \bs p^2 + m^2.
\end{eqnarray*}

Now we recall the expression for the spin in the quasi-classical limit, which follows from the Volkov solution of the Dirac equation:
\begin{eqnarray*}
\psi_p(\bs r, t) = \left[1+\frac{e}{2(k\cd p)}\hat k \hat A\right]\frac{u(p)}{\sqrt{2V\ep_0}}e^{iS},
\end{eqnarray*}
where
\begin{eqnarray*}
S= -p\cd x - \int_0^{k\cd x}{\left[\frac{e}{(k\cd p)}\fp{p}{A} - \frac{e^2}{2\fp{k}{p}}A^2\right]d\phi}.
\end{eqnarray*}
Here $A$ is the four-potential of the classical field, and $u(p)$ the constant bispinor which coincides with that for the free electron. The calculation of the spin using definition (\ref{29})  with the use of the density matrix (\ref{30}) yields
\begin{eqnarray}\label{32}
\langle\bs s\rangle = \bs a\frac{m}{\ep} +\frac{me}{\ep\fp{k}{p}}\left(\bs k\fp{A}{a}-\bs A\fp{k}{a}\right) \nonumber
\\
- \frac{me^2}{2\ep\fp{k}{p}^2}\bs k \fp{k}{a}A^2.
\end{eqnarray}
The time dependence in (\ref{32}) is contained only in the four-potential $A$, which is a periodic function (\ref{17}). In this case, the components of the electron spin include parts which oscillate at the frequency of the classical field. If one averages the expression (\ref{32}), the linear terms in $A$ vanish and the mean value becomes
\begin{align}\label{33}
\langle \bar{\bs s}\rangle = \bs a\frac{m}{\ep} - \frac{me^2}{2\ep\fp{k}{p}^2}\bs k \fp{k}{a}\bar {A^2},
\end{align}
where the bar on the top of the variable denotes averaging over the initial phases of the electron in the beam when entering into the area of space with the field.

As follows from Eq. (\ref{33}), the observation of the change in spin dynamics caused by the influence of the field is possible only if the amplitude $A_0$ of the four-potential is comparable with the electron energy $\ep$.
This amplitude is connected with the average number of photons through
\begin{equation*}
A_0=\frac{\sqrt{2\bar{n}}}{\sqrt{V\om}}.
\end{equation*}
This means that the field quantum number $\bar{n}$ should be large, and this corresponds to the limits (\ref{24}).

Let us come back to the quantum case. One should insert the linear combination of wave functions (\ref{20}) with the coefficients (\ref{22}) into the spin definition (\ref{29}). This leads to
\begin{eqnarray}\label{34}
\langle s^{\mu} \rangle &=&\frac{1}{\sum_{n^{\prime\prime}}\int d\bs q^{\prime\prime}|C_{\bs q^{\prime\prime},n^{\prime\prime}}|^2} \int  d\bs q d\bs q^\prime \sum_{n^\prime}\sum_{n}  e^{i(q^\prime - q)x} \nonumber
\\
&\cdot&\langle n^\prime| S^\dag_{q^\prime}\frac{\bar u(p_{n^\prime})}{\sqrt{2\ep_{n^\prime}}} \Big[1+\frac{(a+a^\dag)}{2\fp{q^{\prime}}{k}}\hat b \hat k\Big]\gamma^5 \gamma^\mu
\\
&\cdot&\Big[1+\frac{(a+a^\dag)}{2\fp{q}{k}}\hat k \hat b\Big]\frac{u(p_{n})}{\sqrt{2\ep_n}}
S_q |n\rangle C^*_{\bs q^{\prime} , n^\prime}C_{\bs q,n}. \nonumber
\end{eqnarray}

We start the evaluation of (\ref{34}) by calculating the matrix element between the field states
\begin{eqnarray}\label{35}
T_{n^\prime n} &=& \langle n^\prime| S^\dag_{q^\prime}\frac{\bar u(p_{n^\prime})}{\sqrt{2\ep_{n^\prime}}} \Big[1+\frac{(a+a^\dag)}{2\fp{q^\prime}{k}}\hat b \hat k\Big]\gamma^5 \gamma^\mu \nonumber
\\
&\cdot&\Big[1+\frac{(a+a^\dag)}{2\fp{q}{k}}\hat k \hat b\Big]\frac{u(p_{n})}{\sqrt{2\ep_n}}S_q|n\rangle,
\end{eqnarray}
or, expanding the brackets,
\begin{eqnarray}\label{36}
T_{n^\prime n}& = &\gamma^5\gamma^\mu\langle n^\prime|S^\dag_{q^\prime}S_q|n\rangle 
+\frac{\gamma^5\gamma^\mu\hat k \hat b}{2\fp{q}{k}}[ \sqrt{\varkappa}\langle n^\prime|S^\dag_{q^\prime}S_q(a+a^\dag)|n\rangle \nonumber
\\
&+&2\alpha \langle n^\prime|S^\dag_{q^\prime}S_q|n\rangle] +\frac{ \hat b \hat k \gamma^5\gamma^\mu}{2\fp{q^\prime}{k}}[ \sqrt{\varkappa^\prime}\langle n^\prime|(a+a^\dag)\nonumber
\\
&\times& S^\dag_{q^\prime}S_q|n\rangle+2\alpha^\prime \langle n^\prime|S^\dag_{q^\prime}S_q|n\rangle] +\frac{\hat b \hat k \gamma^5\gamma^\mu\hat k \hat b}{4\fp{q^\prime}{k}{\fp{q}{k}}}\nonumber
\\
&\times&(\sqrt{\varkappa^\prime}\sqrt{\varkappa}\langle n^\prime|(a+a^\dag)S^\dag_{q^\prime}S_q(a+a^\dag)|n\rangle \nonumber
\\
&+&4\alpha\alpha^\prime\langle n^\prime|S^\dag_{q^\prime}S_q|n\rangle
+2\alpha\sqrt{\varkappa^\prime}\langle n^\prime|(a+a^\dag)S^\dag_{q^\prime}S_q|n\rangle \nonumber
\\
&+&2\alpha^\prime\sqrt{\varkappa}\langle n^\prime|S^\dag_{q^\prime}S_q(a+a^\dag)|n\rangle).
\end{eqnarray}

The evaluation of the matrix element (\ref{36}) is reduced to the calculation of the matrix element  $\langle n^\prime|S^\dag_{q^\prime}S_q|n\rangle$ with $n \neq n^\prime$ in the general case. This matrix element can be calculated approximately, using the  cumulant expansion method that corresponds to a non-perturbative partial summation of the infinite series \cite{Kubo}. A generalization of this approach for non-diagonal matrix elements is developed in the paper and the details of the calculation are described in Appendix C. The result of the evaluation is
\begin{eqnarray}\label{37}
\langle n^\prime| S^\dag_{q^\prime}S_q|n\rangle&& \nonumber
\\
&=& \delta_{n,n^\prime}e^{-\frac{\delta^2 n}{\vk} - \frac{\Delta^2 n^2}{4}} +  \frac{\delta\sqrt{n}}{\sqrt{\vk}}e^{-\frac{\delta \Delta}{2\sqrt{\vk}}n^{\frac{3}{2}}}(\delta_{n-1,n^\prime}
- \delta_{n+1,n^\prime}) +    \nonumber
\\ 
&+& \frac{ n \Delta  }{2}e^{\frac{\delta^2 n}{2\vk}}( \delta_{n-2,n^\prime}
- \delta_{n+2,n^\prime}),
\end{eqnarray}
where $ \delta_{n,n^\prime}$ is the Kroneker symbol, $\delta = \alpha_{q^\prime}-\alpha_q$, and $\Delta = \eta_{q^\prime} - \eta_q$. The matrix elements with additional creation and annihilation operators in (\ref{36}), for example, $\langle n^\prime|S^\dag_{q^\prime}S_q(a+a^\dag)|n\rangle$, can be obtained from Eq. (\ref{37}) by shifting indices, multiplying by the corresponding power of $n$, and leaving the terms in which $n$ changes by no more than two. 

Insertion of Eq. (\ref{37}) into Eq. (\ref{36}) with the use of the approximation (\ref{25}) for the parameters leads to
\begin{widetext}
\begin{eqnarray}\label{38}
\langle s^\mu\rangle &=& \frac{1}{\sum_n\int d\bs q |C_{\bs q,n}|^2}\int  d\bs q d\bs q^\prime \sum_{n}  e^{i(q^\prime - q)x}\frac{\bar u(p_{n^\prime})}{\sqrt{2\ep_{n^\prime}}} \Bigg\{ e^{-\delta^2 n - \frac{\Delta^2 n^2}{4}}\left(\gamma^5\gamma^\mu+\frac{\hat b \hat k \gamma^5\gamma^\mu\hat k \hat b}{2\fp{q^\prime}{k}{\fp{q}{k}}}n\right)C^*_{\bs q^{\prime} , n}C_{\bs q,n} \nonumber
\\
&+&C^*_{\bs q^{\prime} , n-1}C_{\bs q,n}\Bigg[\delta\sqrt{n}e^{-\frac{\delta\Delta}{2}n^{\frac{3}{2}}}\left(\gamma^5\gamma^\mu+\frac{\hat b \hat k \gamma^5\gamma^\mu\hat k \hat b}{2\fp{q^\prime}{k}{\fp{q}{k}}}n\right)
+\left(e^{-\delta^2 n - \frac{\Delta^2 n^2}{4}}+\frac{\Delta n}{2}e^{\frac{\delta^2 n}{2}}\right)\left(\frac{ \hat b \hat k \gamma^5\gamma^\mu}{2\fp{q^\prime}{k}}+\frac{\gamma^5\gamma^\mu\hat k \hat b}{2\fp{q}{k}}\right)\sqrt{n}\Bigg]\nonumber
\\
&+&C^*_{\bs q^{\prime} , n+1}C_{\bs q,n}\Bigg[-\delta\sqrt{n}e^{-\frac{\delta\Delta}{2}n^{\frac{3}{2}}}\left(\gamma^5\gamma^\mu+\frac{\hat b \hat k \gamma^5\gamma^\mu\hat k \hat b}{2\fp{q^\prime}{k}{\fp{q}{k}}}n\right)
+\left(e^{-\delta^2 n - \frac{\Delta^2 n^2}{4}}-\frac{\Delta n}{2}e^{\frac{\delta^2 n}{2}}\right)\left(\frac{ \hat b \hat k \gamma^5\gamma^\mu}{2\fp{q^\prime}{k}}+\frac{\gamma^5\gamma^\mu\hat k \hat b}{2\fp{q}{k}}\right)\sqrt{n}\Bigg]\nonumber
\\
&+&C^*_{\bs q^{\prime} , n-2}C_{\bs q,n}\Bigg[\left(e^{-\delta^2 n - \frac{\Delta^2 n^2}{4}}+\frac{\Delta n}{2}e^{\frac{\delta^2 n}{2}}\right)\frac{\hat b \hat k \gamma^5\gamma^\mu\hat k \hat b}{4\fp{q^\prime}{k}{\fp{q}{k}}}n+\frac{\Delta n}{2}e^{\frac{\delta^2 n}{2}}\gamma^5\gamma^\mu+\delta\sqrt{n}e^{-\frac{\delta\Delta}{2}n^{\frac{3}{2}}}\left(\frac{ \hat b \hat k \gamma^5\gamma^\mu}{2\fp{q^\prime}{k}}+\frac{\gamma^5\gamma^\mu\hat k \hat b}{2\fp{q}{k}}\right)\sqrt{n}\Bigg]\nonumber
\\
&+&C^*_{\bs q^{\prime} , n+2}C_{\bs q,n}\Bigg[\left(e^{-\delta^2 n - \frac{\Delta^2 n^2}{4}}-\frac{\Delta n}{2}e^{\frac{\delta^2 n}{2}}\right)\frac{\hat b \hat k \gamma^5\gamma^\mu\hat k \hat b}{4\fp{q^\prime}{k}{\fp{q}{k}}}n-\frac{\Delta n}{2}e^{\frac{\delta^2 n}{2}}\gamma^5\gamma^\mu\nonumber
\\
&-&\delta\sqrt{n}e^{-\frac{\delta\Delta}{2}n^{\frac{3}{2}}}\left(\frac{ \hat b \hat k \gamma^5\gamma^\mu}{2\fp{q^\prime}{k}}+\frac{\gamma^5\gamma^\mu\hat k \hat b}{2\fp{q}{k}}\right)\sqrt{n}\Bigg]\Bigg\}\frac{u(p_n)}{\sqrt{2\ep_n}}.
\end{eqnarray}
Further simplifications  of (\ref{38}) are possible in the limit of large $n$, when the coefficients $C_{\bs q,n}$ (\ref{22}) can be estimated using the asymptotics  of the Hermitian polynomials and Stirling's formula for the factorial, i. e.
\begin{eqnarray}\label{39}
\lim_{n\rightarrow \infty , \, x\rightarrow \infty }H_n(x)\rightarrow 2^n x^n,\quad n! \sim_{n \to \infty} \sqrt{2\pi}n^{n+1/2}e^{-n}.
\end{eqnarray}
Insertion of the approximation (\ref{39}) into the coefficients $C_{\bs q,n}$ then yields
\begin{eqnarray}\label{40}
C_{\bs q,n} \approx \frac{(2\pi)^{-1/4}}{(2\pi)^3}\Bigg(\frac{\bar u(p_n)}{\sqrt{2\ep_n}} \gamma^0 \frac{u(p_0)}{\sqrt{2\ep_0}}\nonumber
+\frac{\bar u(p_n)}{\sqrt{2\ep_n}} \hat b \hat k \gamma^0 \frac{u(p_0)}{\sqrt{2\ep_0}}\frac{\sqrt{n}}{2\fp{q}{k}}\Bigg)M_{\bs q, n} ,\nonumber
\\
M_{\bs q, n} = \int d\bs r e^{-i(\bs q - \bs p_0)(\bs r - \bs r_0) - |\theta|^2/2+\alpha\theta - \alpha^2/2 +\frac{1}{2}(\theta - \alpha)^2 \eta - \frac{1}{2}(n+\frac{1}{2})\ln n +\frac{n}{2}+n\ln(\theta - \alpha)}.
\end{eqnarray}
Equation (\ref{38}) contains various products of the coefficients $C_{\bs q,n}$ and the complex conjugate $C^{*}_{\bs q, n^{\prime}}$, involving various combinations of  $n$ and $n^{\prime} $. These products can be written in general form as
\begin{equation}\label{41}
e^{i(q^\prime - q)\cdot x}C^*_{\bs q^{\prime} , n+l}C_{\bs q,n}
= \frac{(2\pi)^{-\frac{1}{2}}}{(2\pi)^6}A^*_{q^\prime}A_q\int d\bs r^\prime d\bs r e^{\Phi_l(t, \bs x,\bs q, \bs q^\prime, \bs r , \bs r^\prime,n)},
\end{equation}
where the phase function
\begin{eqnarray}\label{42}
\Phi_l(t, \bs x, \bs q, \bs q^\prime, \bs r , \bs r^\prime,n) = it(q^{0\prime}_{n+ l}-q^0_n)-i(\bs q^\prime - \bs q)\cdot \bs x \nonumber+i(\bs q^\prime - \bs p_0)(\bs r^\prime - \bs r_0) - i(\bs q - \bs p_0)(\bs r - \bs r_0) - \beta^2 \nonumber
\\
+ \beta\left(\alpha^\prime e^{-i\thp{k}{r^\prime}}+\alpha e^{i\thp{k}{r}}\right)
-\left(n+\frac{1}{2}\right)\ln n +n + n\left( \ln \beta^2 -i\bs k(\bs r^\prime - \bs r)-\frac{1}{\beta}\left(\alpha^\prime e^{i\thp{k}{r^\prime}}+\alpha e^{-i\thp{k}{r}}\right)\right) \nonumber
\\
+\frac{\beta^2}{2}\left(\eta^\prime e^{-2i\thp{k}{r^\prime}}+\eta e^{2i\thp{k}{r}}\right)-\frac{l}{2}\ln n + l(\ln \beta -i\thp{k}{r^\prime})
\end{eqnarray}
with $A_q$ being the non-oscillating amplitude factor and index $l \in \{0,1,2\}$.

The evaluation of (\ref{38}) for the average value of the spin will be carried out in two steps. In the first step, we will calculate the integrals over the variables $\bs r$, $\bs r^\prime$, $\bs q$, and $\bs q^\prime$, and sum over the field quantum number $n$. In the second step, we then average the matrix element over a spin subspace.

The integrations over the variables $\bs r$ and $\bs r^\prime$ will be performed in the coordinate system with the $z$ and $z^\prime $ axes directed along the $\bs k$ vector. Therefore, any vectors can be written in the form $\bs y = \bs y_\perp + \bs y_z$, with $\bs y_z$ being directed along $z$ and $z^\prime$ and $\bs y_\perp$ being directed perpendicular to the latter. Then the integrations over $\bs r_\perp$ and $\bs r^\prime_\perp$ give rise to the product of $\delta$-functions $\delta(\bs q^\prime_\perp - \bs p_{0\perp})\delta(\bs q_\perp - \bs p_{0\perp})$, which removes the integration over $\bs q^\prime_\perp$ and $\bs q_\perp$ and leads to the conservation law
\begin{equation}\label{43}
\bs q_\perp = \bs q_\perp^\prime = \bs p_{0\perp}.
\end{equation}
Then the phase function (\ref{42}) transforms to
\begin{eqnarray}\label{44}
\Phi( q_z,   q_z^\prime, z, z^\prime,n) &=& it(q^{0\prime}_{n\pm l}-q^0_n)-i(  q_z^\prime -   q_z) z_{i} + i( q_z^\prime -  p_{0z})z^\prime - i( q_z -   p_{0z}) z - \beta^2 + \beta\left(\alpha^\prime e^{-i\omega z^\prime }+\alpha e^{i\omega z}\right) 	\nonumber
\\
&-&\left(n+\frac{1}{2}\right)\ln n +n+ n\left( \ln \beta^2 -i\omega( z^\prime - z)-\frac{1}{\beta}\left(\alpha^\prime e^{i\omega z^\prime }+\alpha e^{-i\omega z}\right)\right)
+\frac{\beta^2}{2}\left(\eta^\prime e^{-2ikz^\prime}+\eta e^{2ikz}\right)\nonumber
\\
&-& \frac{l}{2}\ln n + l(\ln \beta -i\omega z\prime ), \nonumber
\\
&&q^0_n = \omega n + \sqrt{p_{0\bot}^2 + (q_z - \omega n)^2 + m^2}; \ z_{i} = \frac{\bs x \bs k}{\omega} - z_0.
\end{eqnarray}

The change of variables
\begin{eqnarray*}
	q_z - \omega n \rightarrow q_z;  \  q_z \rightarrow q_z + \omega n.
\end{eqnarray*}
then modifies the phase:
\begin{eqnarray}\label{45}
\Phi( q_z,   q_z^\prime, z, z^\prime,n) &=& i\omega lt + i t \left(\sqrt{p_{0\bot}^2 + (q'_z - \omega l)^2 + m^2}  - \sqrt{p_{0\bot}^2 +  q_z^2 + m^2}\right) -i(  q_z^\prime -   q_z) z_{i}  + i( q_z^\prime z' - q_z z)	\nonumber
\\ 
&+& i (\omega n -  p_{0z})(z^\prime - z)   -\beta^2+ \beta\left(\alpha^\prime e^{-i\omega z^\prime }+\alpha e^{i\omega z}\right) - \left(n+\frac{1}{2}\right)\ln n +n+ n\Bigg( \ln \beta^2 -i\omega( z^\prime - z)\nonumber
\\
&-&\frac{1}{\beta}\left(\alpha^\prime e^{i\omega z^\prime }+\alpha e^{-i\omega z}\Bigg)\right)+\frac{\beta^2}{2}\left(\eta^\prime e^{-2ikz^\prime}+\eta e^{2ikz}\right)- \frac{l}{2}\ln n + l(\ln \beta -i\omega z\prime ).
\end{eqnarray}
Now one should average over the coordinate   $z_{i}$, which corresponds to   averaging over the initial electron coordinate $\bs r_0$ in the uniform electron beam in real experiments. The averaging results in an additional $\delta$-function, $\delta(q_z-q^\prime_z)$, which removes the integration over $q^\prime_z$ and leads to the conservation law
\begin{eqnarray*}
	q_z = q^\prime_z.
\end{eqnarray*}
Let us estimate the values  $\delta$ and $\Delta$ after the integrations have been already performed. Using the definition
\begin{eqnarray}\label{46}
\delta = \alpha_{q^\prime} - \alpha_q = \frac{\partial \alpha}{\partial \bs q}( \bs q^\prime - \bs q) + \frac{\partial \alpha}{\partial n}l, \quad \Delta = \eta_{q^\prime} - \eta_{q} = \frac{\partial \eta}{\partial \bs q} ( \bs q^\prime - \bs q) + \frac{\partial \eta}{\partial n}l,
\end{eqnarray}
these values are equal to zero within the considered accuracy because $\bs q^\prime = \bs q$ and the derivatives over $n$ also vanish, i. e.
\begin{eqnarray}\label{47}
\alpha  =-\frac{\fp{q}{b}}{\fp{q}{k}}= \frac{\thp{q_\perp}{b}}{k(\sqrt{p_{0\perp}^2+m^2+q_z^2}-q_z)}, \quad
\frac{\partial \alpha}{\partial n} = 0, \quad
\eta = \frac{b^2}{2\fp{q}{k}}= \frac{b^2}{2k(\sqrt{p_{0\perp}^2+m^2+q_z^2}-q_z)} , \quad
\frac{\partial \eta}{\partial n} = 0.
\end{eqnarray}
The spin (\ref{38}) then transforms to
\begin{eqnarray}\label{48}
\langle s^\mu(t)\rangle =&& \frac{1}{\sum_n\int d\bs q |C_{\bs q,n}|^2}\int  d q_z d z^\prime dz \sum_{n}|A_q|^2 \Big\{  e^{\Phi_0}\frac{\bar u(p_{n})}{\sqrt{2\ep_{n}}}\gamma^5\gamma^\mu\frac{u(p_n)}{\sqrt{2\ep_n}} + \frac{\bar u(p_{n})}{\sqrt{2\ep_{n}}}\frac{\hat b \hat k \gamma^5\gamma^\mu\hat k \hat b}{4\fp{q}{k}^2}n\frac{u(p_n)}{\sqrt{2\ep_n}}\left(  e^{\Phi_0}+  e^{\Phi_{-2}}+  e^{\Phi_2}\right) \nonumber
\\
&+&\frac{\bar u(p_{n})}{\sqrt{2\ep_{n}}}(\hat b \hat k \gamma^5 \gamma^\mu+\gamma^5\gamma^\mu\hat k \hat b)\frac{\sqrt{n}}{2\fp{q}{k}}\frac{u(p_n)}{\sqrt{2\ep_n}}\left(  e^{\Phi_{-1}}+ e^{\Phi_1}\right)
\Big\}.
\end{eqnarray}

The main contributions to the sum over the field quantum number $n$ arise from values of $n$ near the quasi-classical value $\bar n \gg 1$. This gives the possibility of changing the summation over $n$ to an integration over the complex variable. Then this integral in the complex plane can be evaluated using the same approach as in \cite{Leonov}. This approach is based on the saddle point method \cite{Morse}. Here the saddle point is defined by the first derivative of the phase
\begin{eqnarray}\label{49}
\Phi( q_z,   q_z^\prime, z, z^\prime,n) &=& i\omega t l + i t (\sqrt{p_{0\bot}^2 + (q_z - \omega l)^2 + m^2}  - \sqrt{p_{0\bot}^2 +  q_z^2 + m^2}) + i q_z ( z' -   z) -  ip_{0z}(z^\prime - z) -\beta^2 \nonumber
\\
&+& \beta\left(\alpha^\prime e^{-i\omega z^\prime }+\alpha e^{i\omega z}\right) +\frac{\beta^2}{2}\left(\eta^\prime e^{-2ikz^\prime}+\eta e^{2ikz}\right)-\left(n+\frac{1}{2}\right)\ln n +n \nonumber
\\
&+& n\left( \ln \beta^2-\frac{1}{\beta}\left(\alpha^\prime e^{i\omega z^\prime }+\alpha e^{-i\omega z}\right)\right) \mp \frac{l}{2}\ln n \pm l(\ln \beta -i\omega z\prime ).
\end{eqnarray}

This leads to
\begin{equation}\label{50}
\frac{\partial\Phi}{\partial n} =-\frac{\alpha}{\beta}\left(e^{ikz^\prime}+e^{-ikz}\right) -\ln n + \ln \beta^2 =0.
\end{equation}
The value $\beta = \sqrt{\bar{n}}$ is a large number, and the leading terms in the series in $1/\beta$ lead to the solution $n_0 $ for the saddle point:
\begin{equation}\label{51}
n_0 = \beta^2 -\alpha \beta (\cos kz^\prime+\cos kz) -i\alpha\beta (\sin kz^\prime - \sin kz).
\end{equation}

The zeroth order solution for $n_0$, which is equal to $\beta^2$, yields the quasi-classical limit. The first order corrections, proportional to the parameter $\beta\alpha$, then give rise to the desired quantum effects.

The substitution of Eq. (\ref{51}) into the phase (\ref{49}) gives
\begin{eqnarray}\label{52}
\Phi_l (n_0) = iklt\left(1-\frac{q_z}{\sqrt{p_{0\perp}^2+m^2+q_z^2}}\right)+i(q_z-p_{0z})(z^\prime -z)-2i\alpha\beta (\sin kz^\prime - \sin kz) \nonumber
\\
+\frac{\eta\beta^2}{2}(\cos 2kz^\prime+\cos 2kz)-i\frac{\eta\beta^2}{2}(\sin 2kz^\prime - \sin 2kz)-\ln \beta - ilkz^\prime
\end{eqnarray}
\end{widetext}
and the integrals in the expression for the spin have the form
\begin{equation}\label{53}
\int dn d q_z d z^\prime dz \sum_{n}\Big\{ A_q^*A_q e^{\Phi_l(n_0) + \frac{\Phi^{\prime\prime}_{nn}}{2!}\left(n-n_0\right)^2}\Big\},
\end{equation}
where $\Phi^{\prime\prime}_{nn}$ denotes the second derivative calculated at the saddle point $n_0$.

The phase $\Phi_l (n_0) $ has linear and second order terms in $\beta$. We now show that the terms in the phase $\Phi_l (n_0) $ which are quadratic in the coherent state parameter $\beta$ can be neglected for intensities up to including $10^{18}\ \textrm{W}/\textrm{cm}^2$.\footnote{Note, for intensities higher than $10^{18}\ \textrm{W}/\textrm{cm}^2$ the parameters $\alpha \beta$ and $\eta \beta^2$ are comparable. That is why the inclusion of $\eta \beta^2$ leads to the replacement of Bessel functions by generalized Bessel functions in the sum (\ref{57}). In this case an analytical evaluation of the sum (\ref{57}) is impossible and numerical methods should be used instead.} For this purpose, let us estimate the absolute values of $\beta\alpha$ and $\eta\beta^2$. If we choose the density of photons $\rho = 10^{20}\ \textrm{cm}^{-3}$, the photon frequency $\omega = 10^5 \ \textrm{cm}^{-1}$, and $\gamma = \ep_q/m = 10 $, then the values of the products are
\begin{equation}\label{54}
\beta \alpha \approx e_0 \frac{\sqrt{\rho}\theta}{\sqrt{\omega}\omega (\gamma^{-2} + \theta^2)}\sim 10^3
\end{equation}
and
\begin{eqnarray}\label{55}
\beta^2 \eta \approx e_0^2 \frac{ \rho  }{ \omega^2 m \gamma (\gamma^{-2} + \theta^2)} \sim 10^{-1},
\end{eqnarray}
where $e_0$ is the electronic charge and $\theta$ the angle between $\bs k$ and $\bs p$ (in this case, the Doppler effect has its maximum value) i. e.
\begin{eqnarray*}
	\theta\sim\gamma^{-1} \ll 1.
\end{eqnarray*}

This allows us to neglect the second order terms in comparison with the first order ones in (\ref{52}). It should also be noted that the Gaussian integrals over $(n-n_0)^2$ are reduced to the analogous ones in the denominator of the expression (\ref{48}) and don't affect the spin dynamics.

The integrals over $z$, $z^\prime$, and $q_z$ in the expression (\ref{48}) have the following form
\begin{equation}\label{56}
\int dzdz^\prime dq_z e^{\Phi_l(n_0)},\ l=\{ 0,\pm 1, \pm 2\},
\end{equation}
and can be calculated analytically. The details of the evaluation can be found in Appendix D. The result of the calculation is
\begin{eqnarray}\label{57}
\int dzdz^\prime dq_z e^{\Phi_l} =& (2\pi)^2&\sum_{u=-\infty}^{\infty}e^{iklt\left(1-\frac{p_{0z}+kl+ku}{\sqrt(p_{0\perp}^2+m^2+(p_{0z}+kl+ku)^2)}\right)}\nonumber
\\
&\times& J_{-u}(-2\alpha\beta)J_{u+l}(2\alpha\beta).
\end{eqnarray}

The summation over $u$ in (\ref{57}) can be evaluated via the Euler-Maclaurin formula \cite{WhittakerWatson}, where the sum is replaced by an integral. Then the result of the integration is \cite{Gradshtein}
\begin{eqnarray}\label{58}
\int &dz&dz^\prime dq_z e^{\Phi_l}\nonumber
\\  
&=& (2\pi)^2 (-i)^l e^{iklt\left(1-\frac{2p_{0z}+kl}{2\ep_0}\right)}
\cdot J_l(4\alpha\beta \sin{\frac{k^2lt}{2\ep_0}}).
\end{eqnarray}

The norm of the coefficients $C_{\bs q, n}$ can be calculated in the same way as the calculation of the product of $C$-coefficients. This coincides with the integral of $\int dz dz^\prime dq_z e^{\Phi_0}$ and leads to
\begin{eqnarray}\label{59}
&\sum_n\int d\bs q |C_{\bs q,n}|^2 = (2\pi)^2|A|^2 \int dn e^{-\ln \beta + \frac{\Phi^{\prime\prime}_{nn}}{2!}\left(n-n_0\right)^2}.
\end{eqnarray}
Therefore, when we insert the norm into the spin expression (\ref{48}), it cancels the integral over $n$ and $e^{\Phi_0}$.

The last step is to calculate the average in spin space. For this purpose, the zeroth order value for $n_0$ can be inserted into all pre-exponential terms. The details of the calculation can be found in Appendix E. The calculated value of the spin is
\begin{eqnarray}\label{60}
\langle s^{\mu}(t) \rangle &=
&\frac{m}{\ep_0}a_0^\mu - \frac{m}{\ep_0}k^\mu\fp{a_0}{k}\frac{\beta^2 b^2}{\fp{p_0}{k}^2}\left(1+\mathrm{Re}\,\Pi_2\right) \nonumber
\\
&+&\left[\frac{m}{\ep_0}\frac{\beta}{\fp{p_0}{k}}\left(k^\mu\fp{a_0}{b}-b^\mu\fp{a_0}{k}\right)\right]2\mathrm{Re}\,\Pi_1,
\\
&\Pi_l& = (-i)^l e^{i\om lt\left(1-\frac{p_{0z}}{\ep_0}\right)}J_l(4\alpha\beta \sin{\frac{\om^2lt}{2\ep_0}}), \quad l = 1,2, \nonumber
\end{eqnarray}
where $\ep_0 = \sqrt{\bs p_0^2+m^2}$, $a_0$ is the initial four-vector of the electron spin, $p_0 = (\ep_0, \bs p_0)$, and $J_l$ the Bessel function of order $l$.

For the analysis of the electron polarization, we also need to calculate the electron current density $j^\mu (\bs x, t) = \langle \psi|\gamma^\mu \delta(\bs x  - \bs r^\prime) |\psi\rangle/\langle \psi|\psi\rangle$ averaged 
over the initial coordinates of the electron, $\bs r_0$. This value can be evaluated in the same way as for the electron spin density. Here we present only the final result.
\begin{eqnarray}\label{61}
j^\mu (t) &=& \frac{p^\mu}{\ep_0} - \frac{\beta b^\mu}{\ep_0}2\mathrm{Re}
\Pi_1 +\frac{k^\mu}{\ep_0}\Bigg(\frac{\beta\fp{b}{p}}{\fp{k}{p}}2\mathrm{Re}\Pi_1\nonumber
\\
&-&\frac{\beta^2 b^2}{\fp{k}{p}}\left(1+\mathrm{Re}\,\Pi_2\right)\Bigg)
\end{eqnarray}

Finally, let us prove that the quantum spin vector $\bs s$ coincides with the corresponding quasi-classical Volkov one, when the former is not averaged in time. This corresponds to the quantum case when the averaging over the initial coordinate $z_{i}$ in the phase of the expression (\ref{50}) is not performed. For this purpose, we use the saddle point method to evaluate the sum over the field quantum number $n$ without averaging over $z_{i}$. As before, this leads to the same equation for $n$ (\ref{50}), but in this case we preserve only the leading term in $\beta$, which is equal to
\begin{eqnarray*}
	n_0 = \beta^2,
\end{eqnarray*}
and corresponds to the quasi-classical limit.
This means that the values $\alpha\beta$ and $\eta\beta^2$ are equal to zero.

We now change the variable $q_z^\prime$ to $q^\prime_z - \om l$ in (\ref{45}). This leads to the integrals
\begin{eqnarray}\label{62}
I &\sim& e^{i\om l(t-z_i) - \ln\beta} \cdot\int dz^\prime dq^\prime_z e^{i t \sqrt{p_{0\bot}^2 + q_z^{\prime2} + m^2}-iq_z^\prime z_i+ iq_z^\prime z'- i p_{0z}z^\prime}\nonumber
\\
&\cdot&\int dz dq_z e^{-i t \sqrt{p_{0\bot}^2 + q_z^{2} + m^2}+iq_z z_i- iq_z z+ i p_{0z}z}e^{r(\delta,\Delta)},
\end{eqnarray}
where $e^{r(\delta,\Delta)}$ represents additional terms appearing in the cumulant expansion. They are independent of $z$ and $z^\prime$, and thus are not relevant here.

The integrals over $z$ and $z^\prime$ yield the product of delta functions
$\delta(q_z-p_{0z})\delta(q^\prime_z - p_{0z})$, which cancels the integration over $q_z$ and $q^\prime_z$ and removes $r(\delta,\Delta)$. The average in spin space is carried out analogously to the previous calculation for the derivation of Eq. (\ref{60}).

Therefore, the average value of the spin four-vector can be expressed through the integrals 
\begin{eqnarray*}
	I \sim e^{i\om l(t-z_i)},
\end{eqnarray*}
which exactly leads to the quasi-classical Volkov value without time averaging.

\section{Description of the electron polarization}\label{sec4}
\begin{figure}
  \includegraphics[width=\linewidth]{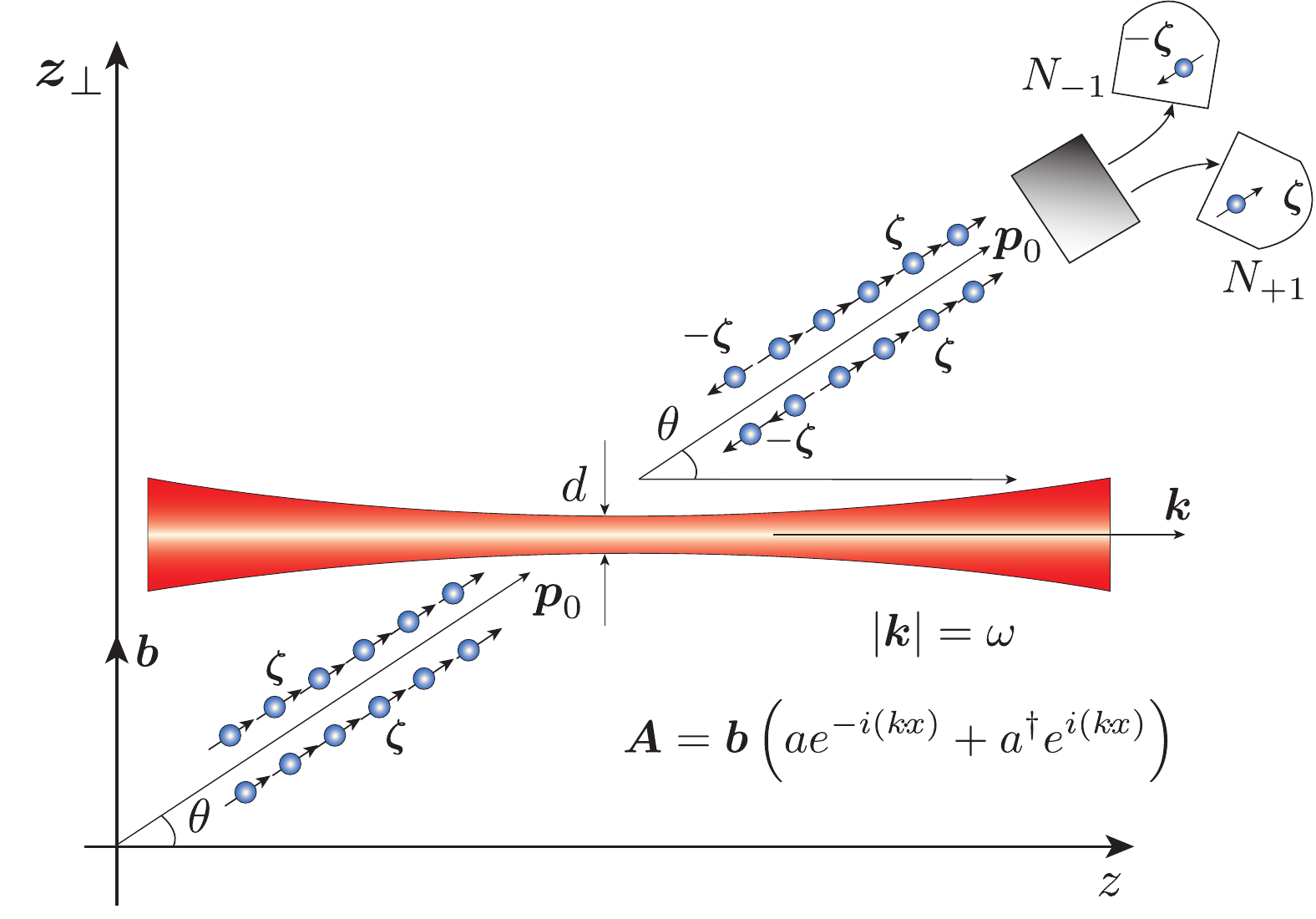}\\
  \caption{Electron motion in a single-mode quantized field. $\bs p_0$ is the initial momentum of the electron, $\bs b$ the polarization vector of the field, the $z$-axis is directed along the wave vector $\bs k$, $a$ and $a^\dag$ are, respectively, the annihilation and creation operators, $\om$ is the frequency of field, and $N_{+1}$ ($N_{-1}$) the number of particles with helicity equal to 1 (-1).}\label{coordinatesystem1}
\end{figure}
The interaction process of the electron with the field can be understood in the following way. For $t<0$, the electron is free, such that it is described by the free solution of the Dirac equation and the field is in a coherent state $\beta$, with the average number of photons $\bar n = \beta^2$.

At $t=0$, the electron crosses the border of the field beam and the interaction starts. We suppose that the boundaries of the field beam are rather sharp. However, the real boundary transition width is not of zero width, and carrying out the above averaging over the initial coordinate $\bs r_0$ corresponds to an averaging over this width. This procedure is widely used in scattering theory and is described in great detail in \cite{Goldberger}. During the interaction, the system of the electron and the field is described by the wave function (\ref{20}). The electron interacts with the field during the time
\begin{eqnarray}\label{63}
t_0 \sim \frac{d}{v_0\sin\theta},
\end{eqnarray}
where $d$ is the ``thickness'' of the laser pulse, $\bs v_0$ the velocity of the electron, and $\theta$ the angle between $\bs k$ and $\bs v_0$ (see Figure \ref{coordinatesystem1}).

At $t = t_0$, the interaction is turned off, the electron becomes free and the detector\footnote{Here we assume that the detector tracks solely the electrons that possess initial momentum $p_0$ while in \cite{Karlovets}, the electrons scattered by the field are tracked (the number of such an electrons is small).} can measure the spin $\bs s(t_0)$. A change in the interaction time $t_0$ will lead to different spin values. This time can be changed in two ways. The first is to change the energy of the electron. The larger the electron energy, the less time it spends in the field. The second way is to change the angle $\theta$. The closer the angle $\theta$ is to $0$ or $\pi$, the more time the electron interacts with the field.

The polarization  of the electron is characterized by its helicity---the projection of the spin $\bs s$ on the direction of its momentum $\bs \nu = \bs p_0/p_0$. The eigenvalues of the helicity operator for a free electron are $\pm1$. According to the general rules of quantum mechanics, the expectation value of the helicity operator can be written as
\begin{eqnarray}\label{64}
\langle\thp{\Sigma}{\nu}\rangle = p_{1}\cdot 1 + p_{-1}\cdot(-1),
\end{eqnarray}
with $p_1$ and $p_{-1}$ being the probabilities of observing the electron with helicity $+1$ and $-1$, respectively. Usually, one considers at the beginning the situation of $p_1$ being unity and $p_{-1}$ zero, and the interest is how $p_1$ and $p_{-1}$ are modified due to the interaction. 

However, in a real experiment, there is no single electron, but rather, a beam of electrons, where $N$ is the number of electrons in the beam. If an initial electron beam was fully polarized, some quantity of electrons with opposite polarization should appear  after the interaction is finished. Namely, the number $N_{-1}$ of such electrons is equal to
\begin{eqnarray}\label{65}
N_{-1} = p_{-1}N
\end{eqnarray}
if at the initial moment of time $t=0$ the helicity of the electron was equal to $+1$.

To calculate the polarization, we choose a coordinate system  with  $z$-axis  directed along $\bs k$. It is further assumed that the initial electron momentum $\bs p_0$ has the angle $\theta$ with  the $z$-axis and that the field is linearly polarized and directed perpendicularly to the $z$-axis (Fig. \ref{coordinatesystem1}). We assume that at the initial moment of time, the vector $\bs\zeta$ (the average spin in the electron's rest frame) is directed along the momentum in the laboratory frame, such that $\bs p\cdot \bs\zeta = p$, which corresponds to the helicity of the electron's being $+1$.

In this coordinate system,
\begin{eqnarray}\label{66}
p_0 &=& (\ep_0 , 0 , p_0\sin\theta, p_0\cos\theta),\quad k = (\om , 0 ,0,\om), \nonumber
\\
b&=&(0,0,b,0),\quad \bs \zeta = (0,\sin\theta,\cos\theta), \nonumber
\\
a_0 &=& \frac{\ep_0}{m}s_0 = \frac{\ep_0}{m}\left(\frac{p_0}{\ep_0},0,\sin\theta,\cos\theta\right). 
\end{eqnarray}
If one uses the definitions (\ref{66}), the four-products in (\ref{60}) can be found as follows:
\begin{eqnarray}\label{67}
\frac{m}{\ep_0}\fp{a_0}{k} &=& \om(v_0-\cos\theta), \quad \frac{m}{\ep_0}\fp{a_0}{b} = -b\sin\theta, \nonumber
\\
\fp{p_0}{k} &=& \om\ep_0(1-v_0\cos\theta), \nonumber
\\
\beta\alpha &=& -\beta \frac{\fp{q}{b}}{\fp{p_0}{k}} = \frac{\beta b}{\ep_0} \frac{\ep_0}{\om}\frac{v_0\sin\theta}{1-v_0\cos\theta},
\end{eqnarray}
where the velocity of the electron is $v_0 = p_0/\ep_0$ and the the total momentum of the system $q = p_0+ k\beta^2$.

In order to find the helicity of the electron after the interaction with the field, we project the spin vector $\bs s$ onto the direction of the electron momentum $\bs \nu = \bs p_0/p_0 = (0,\sin\theta,\cos\theta)$:
\begin{eqnarray}\label{68}
\thp{s}{\nu}= \thp{s_0}{\nu} + \thp{k}{p_0}\fp{s_0}{k}\frac{\beta^2 b^2}{\fp{p_0}{k}^2}\left(1+\textrm{Re}\  \Pi_2\right) \nonumber
\\
+\frac{\beta}{\fp{p_0}{k}}\left(\thp{k}{p}\fp{s_0}{b} - \thp{b}{\nu}\fp{s_0}{k}\right)2\textrm{Re}\  \Pi_1.
\end{eqnarray}
Taking into account that the scalar products are equal to
\begin{eqnarray}\label{69}
\thp{s_0}{\nu} = 1, \quad \thp{k}{\nu} = \om\cos\theta, \quad \thp{b}{\nu} = b\sin\theta
\end{eqnarray}
and inserting them into (\ref{68}), we find
\begin{eqnarray}\label{70}
\thp{s}{\nu} &=& 1+ \frac{\xi^2}{\gamma^2} \frac{(v_0 - \cos\theta)\cos\theta}{(1-v_0\cos\theta)^2}\left(1+\mathrm{Re}\ \Pi_2\right)\nonumber
\\
&-&2\frac{\xi}{\gamma}\frac{v_0\sin\theta}{1-v_0\cos\theta}\mathrm{Re}\  \Pi_1,
\end{eqnarray}
where
\begin{eqnarray*}
\mathrm{Re}\Pi_1 &=& \sin(\om t(1-v_0\cos\theta))\nonumber
\\
&\times& J_1\left(4\frac{\xi}{\gamma}\frac{\ep_0}{\om}\frac{v_0\sin\theta}{1-v_0\cos\theta}\sin\left(\frac{\om}{2\ep_0}\om t\right)\right),
\\
\mathrm{Re}\Pi_2 &=&-\cos\left(2\om t\left(1-v_0\cos\theta\right)\right)
\\
&\times& J_2\left(4\frac{\xi}{\gamma}\frac{\ep_0}{\om}\frac{v_0\sin\theta}{1-v_0\cos\theta}\sin\left(\frac{\om}{\ep_0}\om t\right)\right)
\end{eqnarray*}
and the dimensionless parameter
\begin{equation}\label{71}
\xi= \frac{\beta b}{m}
\end{equation}
was introduced. However, to find the observable quantity which is the polarization, Eq. (\ref{70}) should be normalized by the probability of finding an electron at the observation point, i.e., divide by $j^0 (t)$, the zeroth component of the current density vector, defined in (\ref{61}). Then the observable value of the polarization amounts to
\begin{equation}\label{72}
\frac{\thp{s}{\nu}}{j^0} = \frac{1+\xi^2 f - 2\xi g}{1+\xi^2 f_1 - 2\xi g},
\end{equation}
where 
\begin{eqnarray*}
	f &=& \frac{(v_0 - \cos\theta)\cos\theta}{(1-v_0\cos\theta)^2 \gamma^2}\left(1+\mathrm{Re}\ \Pi_2\right),\\
	g &=& \frac{v_0\sin\theta \mathrm{Re}\  \Pi_1}{(1-v_0\cos\theta)\gamma},\quad
	f_1 = \frac{\left(1+\mathrm{Re}\ \Pi_2\right)}{(1-v_0\cos\theta)\gamma^2}.
\end{eqnarray*}

Eq. (\ref{72}) describes the dependence of the electron polarization   on the interaction time $t$. Using Eqs. (\ref{64}), (\ref{72}), and the condition $p_1+p_{-1} = 1$, the probabilities of finding the electron in the transmitted (non scattered) beam in various polarization states can be calculated as follows
\begin{eqnarray}\label{73}
p_{1} = \frac{1}{2}+\frac{1}{2}\frac{1+\xi^2 f - 2\xi g}{1+\xi^2 f_1 - 2\xi g}, \nonumber
\\
p_{-1} = \frac{1}{2}-\frac{1}{2}\frac{1+\xi^2 f - 2\xi g}{1+\xi^2 f_1 - 2\xi g}.
\end{eqnarray}

These probabilities should be compared with the Volkov probabilities written in the same variables,
\begin{equation}\label{74}
p_{v,1} =\frac{1}{2} +\frac{1}{2} \frac{1+\xi^2f_{v}}{1+\xi^2 f_{1v}}, \quad p_{v,-1} =\frac{1}{2} -\frac{1}{2} \frac{1+\xi^2f_{v}}{1+\xi^2 f_{1v}},
\end{equation}
where $f_v=(v_0 - \cos\theta)\cos\theta/((1-v_0\cos\theta)^2 \gamma^2)$ and $f_{1v}=1/((1-v_0\cos\theta)\gamma^2)$.

One can  see  that the probabilities depend on time in the quantum case, unlike those of Volkov's solution. This means that the quantum fluctuations can change the system dynamics similarly to what takes place for the two level atom. 

Let us find the parameter which governs the slow oscillations in the time evolution in (\ref{73}). For this purpose, we investigate the system evolution for small times $t$. In this case, the sine inside the Bessel functions can be expanded in its Taylor series and one finds that the amplitude of the quantum fluctuations is defined by  the parameter $\xi$ that corresponds to the parameter $x$, which was introduced by Ritus in his work  \cite{Ritus}.

\section{Results and discussion}
\begin{figure}
  \includegraphics[width=\linewidth]{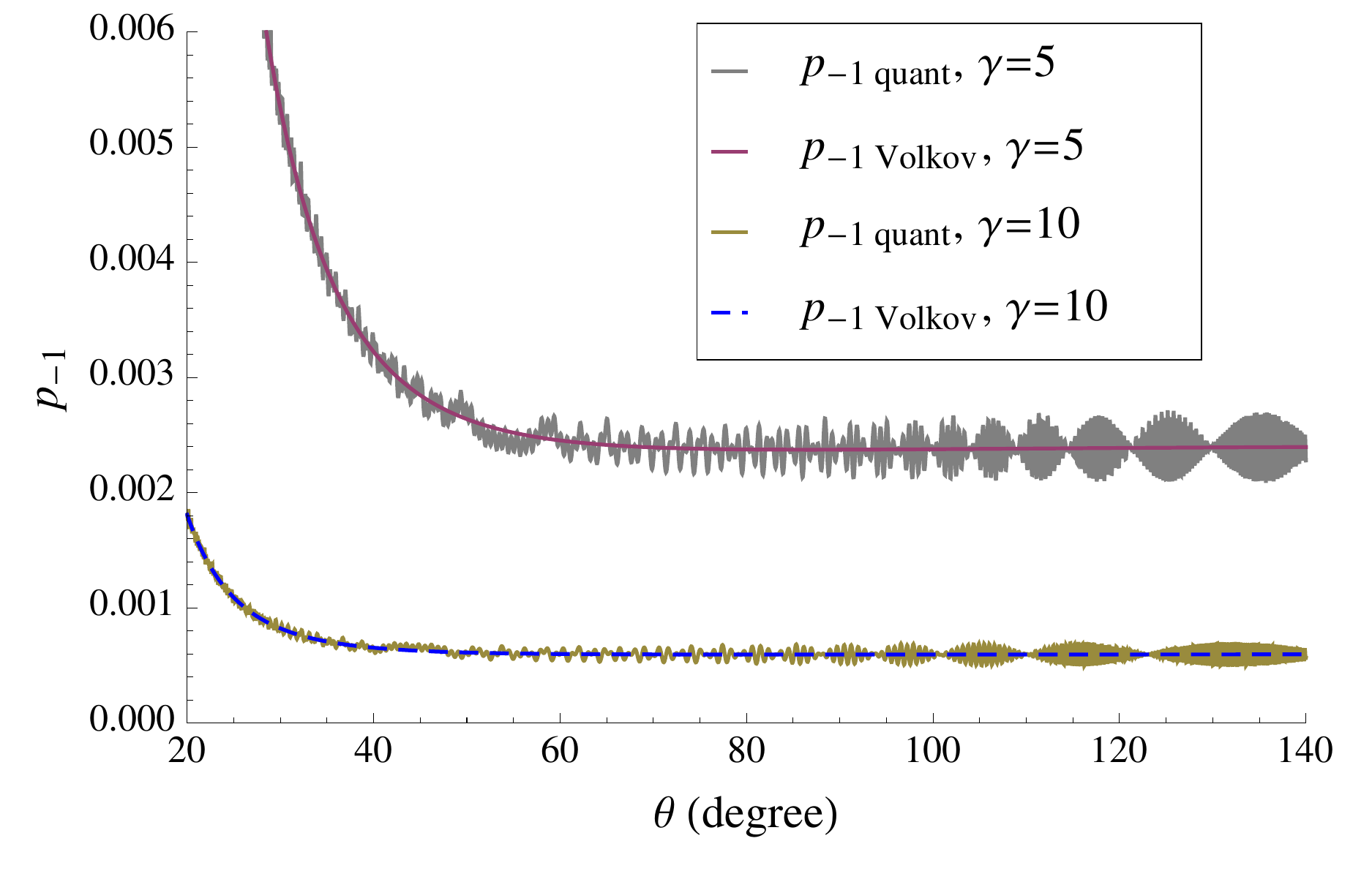}\\
  \caption{(Color online) The probability of finding the electron with flipped polarization as a function of the angle $\theta$ for an intensity $I = 10^{18}\ \textrm{W}/\textrm{cm}^2$, a frequency of the photon $\om = 7.8\cdot10^4\ \textrm{cm}^{-1}$ with a corresponding wavelength of $800$ nm, an initial probability $p_{-1}=0$, and $\gamma$ values of the electron equal to 5 and 10.}\label{fig1}
\end{figure}
Modern lasers can reach nowadays high intensities \cite{PhelixFacility,*Gerstner,*Yanovsky,ELI,*HIPER} up to $10^{22}\  \mathrm{W}/\mathrm{cm}^2$ with a pulse duration of about $30\, \mathrm{fs}$. For our concrete analysis, we choose as intensity $I = 10^{18}\  \textrm{W}/\textrm{cm}^2$, and as photon frequency $\om = 7.8\cdot10^4 \ \textrm{cm}^{-1}$, i. e. a corresponding wavelength of $800$ nm.

As was already mentioned, the interaction time can be adapted either via the electron's energy or by changing the electron's path in the light pulse such as e.g. by varying the entrance angle $\theta$. Fig. \ref{fig1} displays the probability of finding the particles with the flipped polarization in the electron beam as a function of the equal entrance and exit angles $\theta$ between $\bs p_0$ and $\bs k$.
\begin{figure}
  \includegraphics[width=\linewidth]{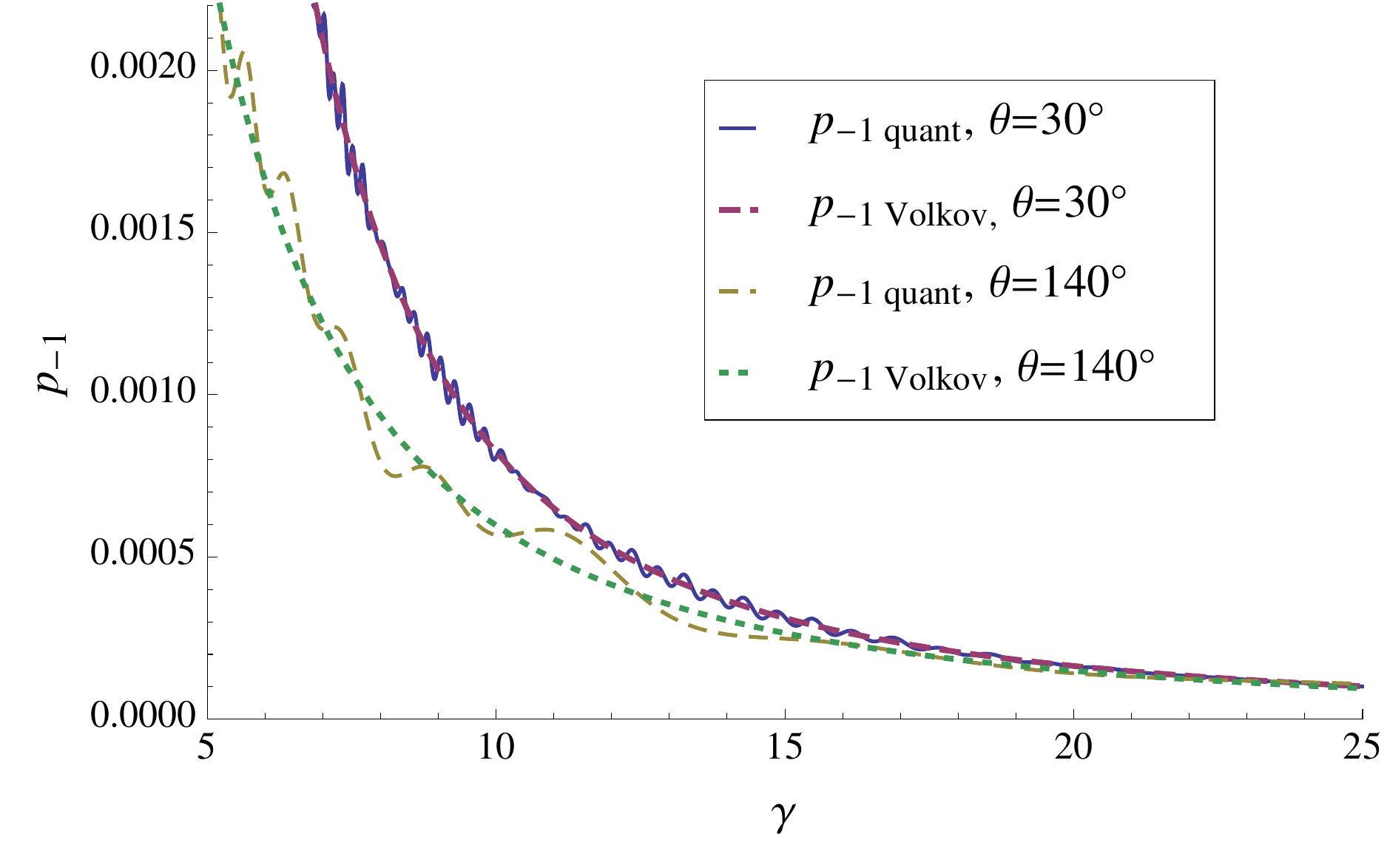}\\
  \caption{(Color online) The probability of finding an electron with flipped polarization as a function of $\gamma$ for an intensity $I = 10^{18}\ \textrm{W}/\textrm{cm}^2$, a frequency of the photon $\om = 7.8\cdot10^4\ \textrm{cm}^{-1}$ with a corresponding wavelength of $800$ nm, an initial probability $p_{-1}=0$, and values of the fly-in angle $\theta$ equal to 30 and 140 degrees.}\label{fig2}
\end{figure}
As can be seen from the graphs, when the interaction time increases, corresponding to larger angles, characteristic structures of the probabilities appear. In addition to fast oscillations at the frequency of field $\om$, there are slow oscillations governed by the parameter $\xi$. These oscillations occur around the mean value, which corresponds to the quasi-classical Volkov case. It should be noted that this special structure appears only when the field is considered as a quantum object: it can not appear in the quasi-classical case.

Fig. \ref{fig2} shows the dependence of the probability of finding an electron with flipped polarization as a function of the electron energy for the two values of fly-in angle $\theta$ equal to 30 and 140 degrees.

\begin{figure}
  \includegraphics[width=\linewidth]{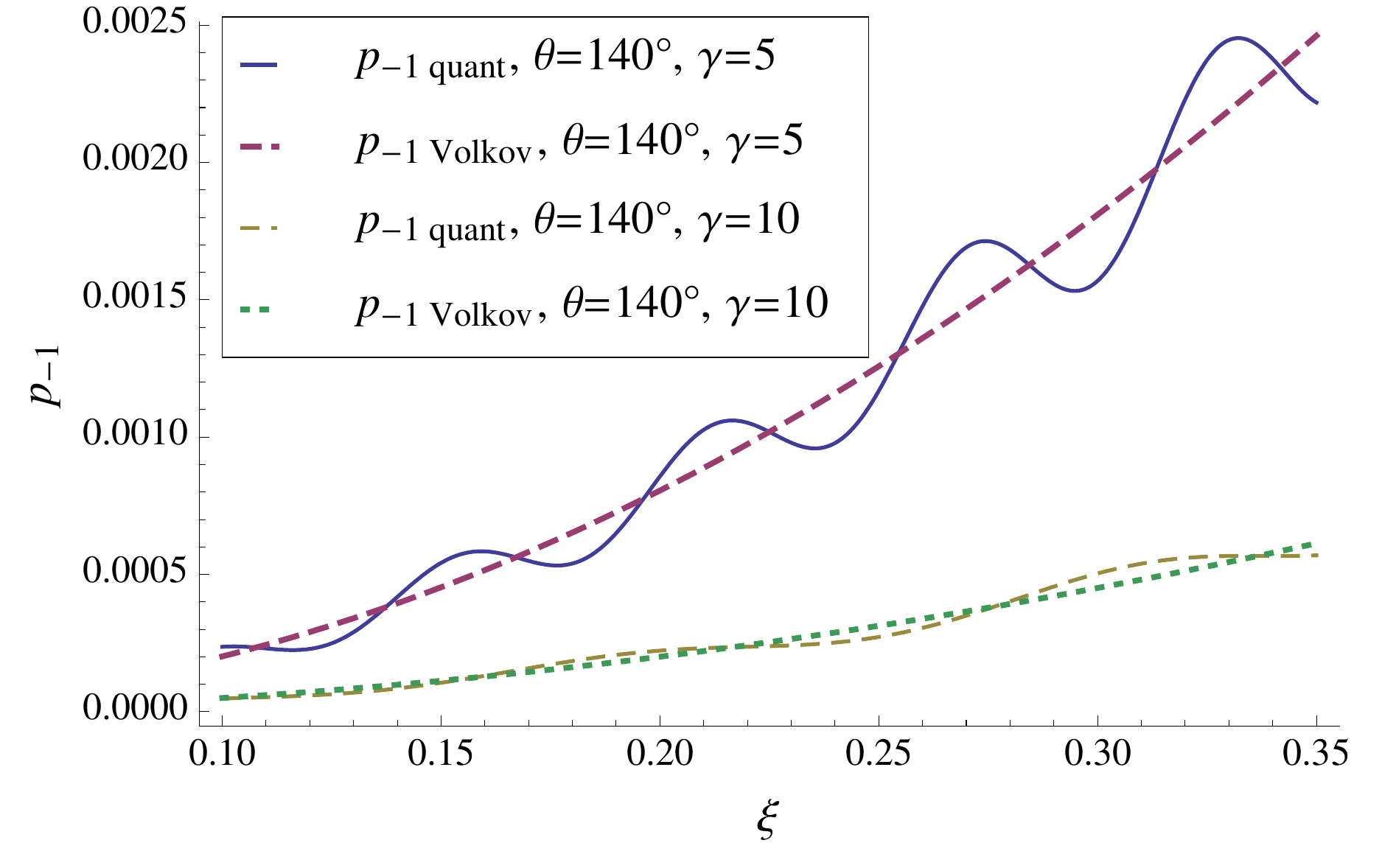}\\
  \caption{(Color online) The probability of finding an electron with flipped polarization as a function of $\xi$ for an entrance angle $\theta$ equal to 140 degrees, a frequency of the photon $\om = 7.8\cdot10^4\ \textrm{cm}^{-1}$ with a corresponding wavelength of $800$ nm, an initial probability $p_{-1}=0$, and values of the $\gamma$ factor of the electron equal to 5 and 10. The parameter $\xi$ equal to $0.1$ corresponds to the intensity $8.4\cdot 10^{16}\ \mathrm{W}/\mathrm{cm}^2$ and $\xi$ equal to $0.35$ to t
As in the previous case, there are  oscillations which appear due to the quantum nature of the electromagnetic field. The average of these oscillations represents the quasi-classical Volkov probability.
he intensity $1.0\cdot10^{18}\ \mathrm{W}/\mathrm{cm}^2$.}\label{fig3}
\end{figure}

Fig. \ref{fig3} shows the dependence of the probability of finding an electron with a flipped polarization as a function of the dimensionless parameter $\xi$, with fly in angle $\theta$ equal to 140 degrees and two values of the gamma factor of the electron equal to 5 and 10. As in Fig. \ref{fig1} and \ref{fig2}, the probability oscillates near the quasi-classical Volkov value. Since probability $p_{-1}$ is inversely proportional to $\gamma$, the spin flip will have larger values for less relativistic electrons. However, $p_{-1}$ is also proportional to $\xi$ and the interaction time $T_\mathrm{int}$ decreases for large $\xi$ and $\gamma$, such that there is trade-off of the various parameters.
In real experiments an electron beam involves a spread in both initial energy and direction. We ensured that an uncertainty of one percent in energy and in the angular distribution does not change the displayed probabilities in above figures visibly. We emphasize from the order of magnitude in the probabilities in Fig. \ref{fig1} \& \ref{fig2} that the number of electrons should be well above 1000 and that mutual interactions shall be avoided with appropriate densities. 

Note that the atomic two-level system possesses an analogous behavior. In this case, the   level population is considered instead of the electron polarization. The fast oscillations in the population inversion essentially depend on the Rabi frequency while for an electron in a quantized field, it can be linked to the oscillations at the frequency of the electromagnetic field $\om$. The structure of the slow oscillations is reminiscent of the ``collapse--revival'' effect for inverted populations, according to which the population inversion of the two-level system vanishes but after some time revives again. This effect is purely quantum mechanical and can not be found in a quasi-classical analysis.

At the end we want to address the influence of the Compton effect on the process considered in our work, because there always is low energy photon emission in which the emerging electron is measured as having unchanged momentum. For that reason the ``momentum unchanged'' channel will be contaminated by Compton electrons. The number of such electrons, estimated via Klein-Nishina formula \cite{Akhiezer} for the employed parameters, does not exceed $10^{-6}$ of the total number of electrons in the beam.

\section{Conclusion}
In this paper we studied the collapse-revival dynamics of an electron in the field of a quantized plane electromagnetic wave in a non-perturbative way, based on the exact solution of the Dirac equation. Important peculiarities were found for the evolution of the electron spin, which has a special structure: fast oscillations at the frequency of the field $\om$, and slow oscillations that correspond to a collapse--revival effect. The slow oscillations are governed by the invariant parameter $\xi = \beta b / m$. This special structure appears due to the quantum nature of the electromagnetic field. In spite the small magnitude of the quantum effects, they can sum up with time and change the system evolution in a measurable way.

The justification of the single-mode approximation is discussed. We prove that when the parameter $\mu = \sqrt{I/\omega^4}$ is much greater then one, the effective single-mode Hamiltonian can be selected, with the help of the Bogolubov canonical transformation.

\section{Acknowledgements}
The authors are grateful to K. Hatsagortsyan, A. Di Piazza, D. Karlovets, S. Cavaletto, S. Meuren and G. G. Paulus for useful discussions.
\section*{Appendix A: Justification of the single-mode approximation}
When the spin oscillations were discussed in Sec. V we considered a realistic laser pulse with a finite duration and width. This means, that in reality we are dealing with a quasi-monochromatic wave packet and not an infinite plane electromagnetic wave. This quasi-monochromatic wave packet has a central frequency $\omega_0$ and a wave vector $\bs k_0 = \omega_0 \bs l$
($\bs l$ is a unit vector). The characteristic spreads in frequency and wave vector (in a solid angle $\Delta \Omega$)
\begin{eqnarray}\label{A1}
	\Delta \omega \sim \frac{1}{\Delta t}, \quad \delta k \approx  \omega_0^2 \Delta \Omega \sim \frac{1}{\Delta S},
\end{eqnarray}
are characterized by the duration $\Delta t$ of the laser pulse and its spatial width $\Delta S$. 

For this reason, the question of validity of the single-mode Hamiltonian (\ref{3}) might arise. Therefore, let us show that for the interaction of the electron and the quasi-monochromatic wave packet an effective single-mode Hamiltonian can be selected.

In order to include in the Dirac equation the interaction with the modes of the field wave packet the summation over the wave vectors belonging to the range $\Delta k = \delta \bs k \Delta \omega$ should be performed. Therefore, in the sum over $k$ the modes which include $\bs k_0$ should be kept. This reduction can be performed by introducing the function $\rho(\bs k- \bs k_0)$ in the sum over $k$ with a sharp maximum near the central wave vector $\bs k_0$. This function determines the form of the wave packet in $k$ space. For this reason, Eq. (\ref{5}) transforms to a form
\begin{eqnarray}\label{A2}
\Bigg(\hat q &-&\sum_{k }  \rho(\bs k - \bs k_0) \hat k a^\dag_k a_k \nonumber
\\
&-& \sum_{ k }\hat b_k \rho(\bs k - \bs k_0) (a_k+a^\dag_k)-m\Bigg)\phi = 0.
\end{eqnarray}

We will characterize the non monochromaticity of the wave packet via two parameters
\begin{eqnarray}\label{A3}
	\chi_{1}&=& \frac{\Delta \omega}{\omega_0} \approx \frac{1}{\omega_0 \Delta t},\quad	\chi_2 = \frac{\delta k}{k_0} \approx \frac{1}{\omega_0\Delta S}\nonumber
	\\
	\chi_1 &\sim& \chi_2 \equiv \chi_k\sim D_k^{-1}.
\end{eqnarray} 
Here $D_k$ is the Q-factor of the laser pulse. The estimation of the Q-factor for the laser parameters used in the previous section gives $D_k \sim 5\cdot 10^2$.

The number of states of an electromagnetic field in a volume of wave vectors $\Delta k$ is large. As a result we can consider them equally populated with the number of photons $\bar n$. Also for simplicity we choose the function $\rho(\bs k - \bs k_0) =1$ in the domain $\Delta k$ and $\rho(\bs k - \bs k_0) = 0$ outside. Such a choice of $\rho(\bs k - \bs k_0)$ corresponds to a plane wave approximation which coincides with the classical Volkov solution in which the four-potential of the field is an arbitrary function of the field phase $\fp{k}{x}$ \cite{LandauQED}. Then the Eq. (\ref{A2}) can be transformed into a form 
\begin{eqnarray}\label{A4}
	H_0\phi = \Bigg(\hat q - \hat k_0 \sum_{\Delta k }  a^\dag_k a_k -  \hat b_0 \sum_{\Delta k }  (a_k+a^\dag_k)-m\Bigg)\phi = 0,
\end{eqnarray}
with $\chi_k$ accuracy. 

We now show that the Eq. (\ref{A4}) can be restricted to the interaction with an intense collective mode of the frequency $\omega_0$ via the method of canonical transformations, which was introduced by Bogolubov and Tyablikov in a polaron theory in the strong field limit \cite{Bogolubov}. For this purpose we go back to the coordinate representation in (\ref{A4})
\begin{eqnarray}\label{A5}
H_0  =   \hat q -  \frac{1}{2}\hat k_0 \sum_{\Delta k }    (p_k^2 + q_k^2) -  \hat b_0 \sqrt{2}\sum_{\Delta k }   q_k -m , \nonumber\\
q_k = \frac{a_k+a^\dag_k}{\sqrt{2}}; \ p_k = - i \frac{\partial}{\partial q_k } = - i \frac{a_k - a^\dag_k}{\sqrt{2}}
\end{eqnarray}
Let us introduce the collective variable $Q$ in which all field modes are added coherently and the ``relative'' field variables $y_k$ which define quantum fluctuations relative to the collective mode
\begin{eqnarray}\label{A6}
Q   =  \sum_{\Delta k }   q_k ; \quad y_k = q_k - \frac{1}{N} Q ;\nonumber\\ q_k = y_k + \frac{1}{N} Q; \quad 
 \sum_{\Delta k }y_k = 0; \quad  N = \sum_{\Delta k}1,
\end{eqnarray}
where $N$ by definition is the number of modes in the range $\Delta k$. Generalized momenta are calculated according to Bogolubov \cite{Bogolubov}
\begin{eqnarray}\label{A7}
  p_k  =  - i \frac{\partial}{\partial q_k }  = - i \left\{  \frac{\partial Q }{\partial q_k }\frac{\partial}{\partial Q} + \sum_{\Delta f}  \frac{\partial y_f }{\partial q_k }\frac{\partial}{\partial y_f}\right\}.
\end{eqnarray}
The calculation of derivatives with the help of (\ref{A6}) yield the generalized momenta
\begin{eqnarray}\label{A8}
 p_k  =  P + p_{y k}; \quad  \sum_{\Delta k } p_{y k} = 0; \nonumber\\
P = - i \frac{\partial}{\partial Q}; \quad p_{y k} =  - i \frac{\partial}{\partial y_k } +   \frac{i}{N} \sum_{\Delta f }  \frac{\partial}{\partial y_f}
\end{eqnarray}
The insertion of (\ref{A8}) and (\ref{A6}) into the Hamiltonian (\ref{A5}) leads then to 
\begin{eqnarray}\label{A9}
H_0  =   \hat q &-&  \frac{1}{2}\hat k_0 \left[ \frac{1}{N}Q^2 + N P^2\right] - \hat b_0 \sqrt{2} \hat Q \nonumber
\\
&-& \frac{1}{2}\hat k_0 \sum_{\Delta k }    (p_{yk}^2 + y_k^2).
\end{eqnarray}
We now quantize the collective and ``relative'' variables by introducing the set of creation and annihilation operators
\begin{eqnarray}\label{A10}
Q =  \frac{\sqrt{N} }{\sqrt{2}} (A + A^\dag); \ P = -i \frac{1 }{\sqrt{2 N}} (A - A^\dag) ;  \nonumber\\
q_{yk}  = \frac{1}{\sqrt{2}}( b_k + b_k^\dag); \ p_{yk}  = - i \frac{1}{\sqrt{2}}( b_k - b_k^\dag).
\end{eqnarray}
The expression for $q_{yk}$ and $p_{yk}$ is valid with accuracy $1/N$. Then the Hamiltonian (\ref{A9}) transforms to the form
\begin{eqnarray}\label{A11}
H_0  =   \hat q - \hat k_0 A^\dag A -  \hat b_0 \sqrt{N}(A + A^\dag)    -  \hat k_0 \sum_{\Delta k }    b_k^\dag b_k,
\end{eqnarray}
where the operators are written in a normal form.

We proceed by estimating the characteristic energies in this Hamiltonian. For this purpose we assume, that the average number of quanta in each mode in coherent state is $\bar n$. Then the average energy of the collective mode
\begin{eqnarray}\label{A12} 
	E_0 \approx \omega_0\langle A^\dag A\rangle \approx \frac{\omega_0}{N}\left(\sum_{\Delta k} \langle a_k\rangle\right)^2 = \omega_0 N \bar n,
\end{eqnarray}
and the average energy of the interaction between the electron and the collective mode is
\begin{eqnarray}\label{A13} 
	E_1\approx b_0 \sum_{\Delta k}\langle a_k\rangle \approx b_0 N \sqrt{\bar n} \approx e_0 N \sqrt{\frac{\bar n}{\omega_0 V}}.
\end{eqnarray}
In order to justify the single-mode approximation we, therefore, should compare $E_1$ with the fluctuations of the energy of the ``relative'' modes
\begin{eqnarray}\label{A14}
	E_f \approx \omega_0\langle\sum_{\Delta k}b^\dag_k b_k\rangle.
\end{eqnarray}
As the energy of the vacuum of the electromagnetic field is not taken into account (operators are in normal form), the average value of $E_f$ is equal to zero. However, the mean square deviation $\delta E_f$ of the noninteracting photon gas is proportional to $\sqrt{N}$ \cite{LandauStat}. For this reason, the ratio of $E_1$ and $\delta E_f$ is characterized via parameter
\begin{eqnarray}\label{A15}
	\mu \equiv \frac{E_1}{\delta E_f} = \sqrt{\frac{\bar n N}{V \omega_0^3}}.
\end{eqnarray} 
This parameter can be estimated using the energy density of the laser pulse
\begin{eqnarray}\label{A16}
	\omega_0 N \bar n = I V.
\end{eqnarray} 
Consequently, the single-mode approximation is valid when the parameter
\begin{eqnarray}\label{A17}
	\mu = \sqrt{\frac{I}{\omega_0^4}}\gg 1,
\end{eqnarray}
which is fulfilled for the chosen range of intensities. For e. g. 800 nm laser pulse the above condition is fulfilled for intensities well above $3.7\cdot 10^6\ \mathrm{W}/\mathrm{cm}^2$.

Concluding, with accuracy $\chi_k$ the effective Hamiltonian can be selected and the problem of the interaction of an electron and quantized field can be described in single-mode approximation. 

\section*{Appendix B: Calculation of the coefficients of the linear combination in Eq. (\ref{22})}
The coefficient $C_{\bs q,n}$ is
\begin{equation}\label{B1}
C_{\bs q,n} = \frac{1}{(2\pi)^3}\int d\bs r \psi^\dag_{qn}e^{i\thp{p_0}{r}}\frac{u(p_0)}{\sqrt{2\ep_0}}|\beta\rangle.
\end{equation}
Inserting the wave function $\psi^\dag_{qn}$ into Eq. (\ref{B1}), one has
\begin{widetext}
\begin{eqnarray}\label{B2}
C_{\bs q n} &=& \frac{1}{(2\pi)^3}\int{ d\bs r e^{-i(\bs q - \bs p_0)\cdot \bs r}}\frac{\bar u(p_n)}{\sqrt{2\ep_n}}\Bigg(\gamma^0\left(1+\frac{2\alpha}{2\fp{q}{k}}\right)\langle n|S^\dag_q e^{-i\thp{k}{r}a^\dag a}e^{\beta a^\dag -\beta^* a}|0\rangle \nonumber
\\
&+&\hat b \hat k \gamma^0 \frac{\sqrt{\varkappa}}{2\fp{q}{k}}\Big(\sqrt{n+1}\langle n+1|S_q^\dag e^{-i\thp{k}{r}a^\dag a}e^{\beta a^\dag -\beta^* a}|0\rangle+\sqrt{n}\langle n-1|S^\dag_q e^{-i\thp{k}{r}a^\dag a}e^{\beta a^\dag -\beta^* a}|0\rangle\Big)\Bigg)\frac{u(p_0)}{\sqrt{2\ep_0}},
\end{eqnarray}
\end{widetext}
where
\begin{eqnarray}\label{B3}
S^\dag_q &=& e^{\frac{\eta_q}{2}(a^2-a^{\dag^2})}e^{-\alpha_q(a^\dag - a)},\quad \varkappa =1/\sqrt{1-2b^2/\fp{q}{k}} , \nonumber
\\
\alpha_q &=& -\fp{q}{b}/\fp{q}{k}\frac{1}{1-2b^2/\fp{q}{k}},
\end{eqnarray}
and coherent state $|\beta\rangle = e^{\beta a^\dag -\beta^* a}|0\rangle$. The index $q$ of the quantities $\alpha_q$ and $\eta_q$ indicates their dependence on $q$.

The problem of calculating the coefficients $C_{\bs q,n}$ reduces to that of calculating a matrix element of the type
 \[\langle n|S^\dag_q e^{-i\thp{k}{r}a^\dag a}e^{\beta a^\dag -\beta^* a}|0\rangle.\]
For this, we need the representation of the exponential of a sum of operators in normal form. The normal form of the operator of the coherent state is \cite{Scully}
\begin{equation}\label{B4}
e^{\beta a^\dag - \beta^* a} = e^{-|\beta|^2/2}e^{\beta a^\dag}e^{-\beta^* a} = e^{|\beta|^2 /2}e^{-\beta^* a}e^{\beta a^\dag}.
\end{equation}
The decomposition of the exponent with the second power of the creation and annihilation operators is
\begin{equation}\label{B5}
e^{\frac{\eta}{2}(a^2-a^{\dag 2})} = e^{-\frac{1}{2}\tah{\eta} a^{\dag2}}e^{-\ln\ch \eta (a^\dag a +\frac{1}{2})}e^{\frac{1}{2}\tah \eta a^{2}}.
\end{equation}
This decomposition is possible since the three operators $a^2$, $a^{\dag2}$ and $a^\dag a$ form a closed algebra. Taking into account the transformation of the creation and annihilation operators with an operator $S^\dag$,
\begin{eqnarray}\label{B6}
S^\dag (a+a^\dag) S = \sqrt{\vk}(a+a^\dag)+2\alpha,
\end{eqnarray}
the action of the creation operator $a^\dag$ on the left bra vector $\langle 0|$ yields zero and the harmonic oscillator state vector is connected with the vacuum by $\langle n| = \langle 0|\frac{a^n}{\sqrt{n!}}$, and  we obtain

\begin{widetext}
\begin{eqnarray}\label{B7}
\langle n|e^{\frac{\eta}{2}(a^2-a^{\dag 2})}e^{-\alpha(a^\dag - a)}e^{i\thp{k}{r}a^\dag a}e^{\beta a^\dag - \beta^* a}|0\rangle \nonumber
\\
=\frac{1}{\sqrt{\textrm{ch}\eta}}e^{-|\theta|^2/2 + \alpha \theta -\alpha^2/2+1/2(\theta -\alpha)^2 \textrm{th}\eta}\langle 0| \frac{((a-\textrm{th}\eta a^\dag)+\frac{\theta - \alpha}{\textrm{ch}\eta})^n}{\sqrt{n!}}|0\rangle.
\end{eqnarray}
In order to calculate the vacuum average in (\ref{B7}), we replace the power $n$ by the $n$th derivative of the exponent
\begin{eqnarray}\label{B8}
\langle 0| \frac{((a-\textrm{th}\eta a^\dag)+\frac{\theta - \alpha}{\textrm{ch}\eta})^n}{\sqrt{n!}}|0\rangle =
\frac{1}{\sqrt{n!}}\frac{d^n}{dx^n}\langle0|e^{x(a-\textrm{th}\eta a^\dag)+x\frac{\theta-\alpha}{\textrm{ch}\eta}}|0\rangle\Big|_{x=0} =
\\
\frac{1}{\sqrt{n!}}\frac{d^n}{dx^n}e^{x\frac{\theta-\alpha}{\textrm{ch}\eta}}\langle0|e^{x(a-\textrm{th}\eta a^\dag)}|0\rangle\Big|_{x=0} =
\frac{1}{\sqrt{n!}}\frac{d^n}{dx^n}e^{x\frac{\theta-\alpha}{\textrm{ch}\eta}-x^2\textrm{th}\eta/2 }\Big|_{x=0}.
\end{eqnarray}
Selecting the full square of the variable $x$ and changing variables, we obtain
\begin{eqnarray}\label{B9}
\frac{1}{\sqrt{n!}}\frac{d^n}{dx^n}e^{x\frac{\theta-\alpha}{\textrm{ch}\eta}-x^2\textrm{th}\eta/2 }\Big|_{x=0} = \frac{1}{\sqrt{n!}}\frac{d^n}{dx^n} e^{-[(\sqrt{\frac{\textrm{th}\eta}{2}}x-\frac{\theta-\alpha}{\sqrt{2\textrm{th}\eta}\textrm{ch}\eta})^2-\frac{(\theta-\alpha)^2}{\textrm{sh}2\eta}]}\Big|_{x=0}, \nonumber
\\
y = \sqrt{\frac{\textrm{th}\eta}{2}}x-\frac{\theta-\alpha}{\sqrt{2\textrm{th}\eta}\textrm{ch}\eta}, \quad
x=0, \quad y = -\frac{\theta-\alpha}{\sqrt{2\textrm{th}\eta}\textrm{ch}\eta},\quad \frac{d^n}{dx^n} = \left(\frac{\textrm{th}\eta}{2}\right)^{\frac{n}{2}}\frac{d^n}{dy^n},
\\
\frac{1}{\sqrt{n!}}\frac{d^n}{dx^n} e^{-[(\sqrt{\frac{\textrm{th}\eta}{2}}x-\frac{\theta-\alpha}{\sqrt{2\textrm{th}\eta}\textrm{ch}\eta})^2-\frac{(\theta-\alpha)^2}{\textrm{sh}2\eta}]}\Big|_{x=0} =  \frac{e^{\frac{(\theta - \alpha)^2}{\textrm{sh}2\eta}}}{\sqrt{n!}}\left(\frac{\textrm{th}\eta}{2}\right)^{\frac{n}{2}}\frac{d^n}{dy^n}e^{-y^2}\Big|_{y = -\frac{\theta-\alpha}{\sqrt{2\textrm{th}\eta}\textrm{ch}\eta}}.
\end{eqnarray}

Using the definition of Hermitian polynomials,
\[H_n(y) = (-1)^n e^{y^2}\frac{d^n}{dy^n}e^{-y^2},\]
we finally obtain
\begin{eqnarray}\label{B10}
\langle n|S^\dag e^{-i\thp{k}{r}a^\dag a}e^{\beta a^\dag -\beta^* a}|0\rangle =
\frac{e^{-|\theta|^2/2 + \alpha \theta -\alpha^2/2+1/2(\theta -\alpha)^2 \textrm{th}\eta}}{\sqrt{\textrm{ch}\eta}}  \frac{1}{\sqrt{n!}}\left(\frac{\textrm{th}\eta}{2}\right)^{\frac{n}{2}} H_n\left(\frac{\theta-\alpha}{\sqrt{2\textrm{th}\eta}\textrm{ch}\eta}\right).
\end{eqnarray}
\end{widetext}
\section*{Appendix C: Calculation of the matrix element $\langle n^\prime|S^\dag_{q^\prime}S_q|n\rangle$ by the cumulant method}

The problem is to calculate the matrix element $\langle n^\prime|S^\dag_{q^\prime}S_q|n\rangle$. Using the transformation by $S$ of the creation and annihilation operators,
\begin{eqnarray}\label{C1}
S_q^\dag (a+a^\dag) S_q =\sqrt{\varkappa}(a+a^\dag)+2\alpha,
\end{eqnarray}
we have
\begin{eqnarray}\label{C2}
\langle n^\prime|S^\dag_{q^\prime}S_q|n\rangle = \langle n^\prime| e^{\frac{\Delta}{2}(a^2-a^{\dag2})}e^{-\frac{\delta}{\sqrt{\varkappa}}(a^\dag -a)}|n\rangle,
\end{eqnarray}
where $\Delta = \eta_{q^\prime}-\eta_q$ and $\delta = \alpha_{q^\prime}-\alpha_q$. The quantities $\Delta$ and $\delta$ have the small parameter $q^\prime - q$. In order to take into account quantum transitions, we investigate only the closest transitions to the diagonal. These transitions are described via the cumulant method:
\begin{eqnarray}\label{C3}
<n'|e^{\lambda A}|n> &=& \delta_{nn'} e^{\sum_{k=1} \lambda^k K_{0k}} + \lambda B_{\pm} \delta_{(n\pm 1)n'} e^{\sum_k \lambda^k K_{\pm 1 k}} \nonumber
\\
&+& \lambda C_{\pm} \delta_{(n\pm 2)n'} e^{\sum_k \lambda^k K_{\pm 2 k}},
\end{eqnarray}
where lambda is the small parameter. If we decompose the left and right hand sides of Eq. (\ref{C3}) and equate terms with the same powers in lambda, the cumulant parameters can be found as:
\begin{eqnarray}\label{C4}
K_{01} &=& 0; \  K_{02} = \frac{1}{2} <n| A^2 |n> ; \nonumber\\
B_{\pm} &=& <n\pm 1|A|n>; \ K_{11} = \frac{1}{2}<n\pm 1|A^2|n>; \nonumber\\
C_{\pm} &=& <n\pm 2|A|n>; \ K_{21} = \frac{1}{2}<n\pm 2|A^2|n>.
\end{eqnarray}
Hence, the matrix element with second order accuracy in parameter $\lambda$ is
\begin{eqnarray}\label{C5}
\langle'|e^{  A}|n\rangle &\approx&  \delta_{nn'} e^{\frac{1}{2} \langle n| A^2 |n\rangle} + \langle n+1|A|n\rangle \delta_{(n+ 1)n'} e^{\frac{1}{2}\langle n+1|A^2|n\rangle}\nonumber
\\
&+& \langle n-1|A|n\rangle \delta_{(n- 1)n'} e^{\frac{1}{2}\langle n-1|A^2|n\rangle} \nonumber
\\
&+& \langle n+2|A|n\rangle \delta_{(n+ 2)n'} e^{\frac{1}{2}\langle n+2|A^2|n\rangle}\nonumber
 \\
&+&\langle n-2|A|n\rangle \delta_{(n- 2)n'} e^{\frac{1}{2}\langle n-2|A^2|n\rangle}.
\end{eqnarray}
In order to apply this method to the operator $e^{\frac{\Delta}{2}(a^2-a^{\dag2})}e^{-\frac{\delta}{\sqrt{\varkappa}}(a^\dag -a)}$, we need to represent the multiplication of two exponentials by one exponential. Let us calculate the commutator of the two operators
\begin{eqnarray}\label{C6}
\left[\frac{\Delta}{2}(a^2-a^{\dag2}),-\frac{\delta}{\sqrt{\varkappa}}(a^\dag-a)\right] = \frac{\Delta\delta}{\sqrt{\varkappa}}(a^\dag-a),
\end{eqnarray}
and therefore the commutators of the type
\begin{eqnarray*}
\Bigg[\Bigg[\Bigg[&&\frac{\Delta}{2}(a^2-a^{\dag2}),-\frac{\delta}{\sqrt{\varkappa}}(a^\dag-a)\Bigg],\frac{\Delta}{2}(a^2-a^{\dag2})\Bigg],
\\
&&...,\frac{\Delta}{2}(a^2-a^{\dag2})\Bigg],
\end{eqnarray*}
are not zero. For this reason, the representation is
\begin{eqnarray}\label{C7}
e^{-\frac{\delta}{\sqrt{\varkappa}}(a^\dag-a)+\frac{\Delta}{2}(a^2-a^{\dag2})}&&\nonumber
\\
 &=&e^{\frac{\Delta}{2}(a^2-a^{\dag2})}e^{-\frac{\delta}{\sqrt{\varkappa}}(a^\dag-a)}e^{-\frac{\delta\Delta}{\sqrt{2\varkappa}}(a^\dag-a)}.
\end{eqnarray}
Calculating the cumulant coefficients using (\ref{C5}) gives
\begin{eqnarray}\label{C8}
\langle n|A^2|n\rangle &=& -\frac{2\delta^2 n}{\vk} - \frac{\Delta^2 n^2}{2},  \nonumber
\\
\langle n\pm 1|A^2|n\rangle &=& -\frac{\delta\Delta}{\sqrt{\vk}}n^{\frac{3}{2}}, \nonumber
\\
\langle n\pm2|A^2|n\rangle &=& \frac{\delta^2 n}{\vk},  \nonumber
\\
\langle n\mp1|A|n\rangle &=& \pm\frac{\delta\sqrt{n}}{\sqrt{\vk}}\left(1-\frac{\Delta}{2}\right), \nonumber
\\
\langle n\mp2|A|n\rangle &=& \pm\frac{\Delta n}{2}.
\end{eqnarray}
Consequently, the unknown matrix element becomes
\begin{eqnarray}\label{C9}
\langle n^\prime| S^\dag_{q^\prime}S_q|n\rangle &=& \delta_{n,n^\prime}e^{-\frac{\delta^2 n}{\vk} - \frac{\Delta^2 n^2}{4}} +\delta_{n-1,n^\prime}\frac{\delta\sqrt{n}}{\sqrt{\vk}}e^{-\frac{\delta\Delta}{2\sqrt{\vk}}n^{\frac{3}{2}}}  \nonumber
\\
&-& \delta_{n+1,n^\prime}\frac{\delta\sqrt{n}}{\sqrt{\vk}}e^{-\frac{\delta\Delta}{2\sqrt{\vk}}n^{\frac{3}{2}}}e+\delta_{n-2,n^\prime}  \nonumber\frac{\Delta n}{2}e^{\frac{\delta^2 n}{2\vk}}
\\
&-& \delta_{n+2,n^\prime}\frac{\Delta n}{2}e^{\frac{\delta^2 n}{2\vk}}.
\end{eqnarray}

\section*{Appendix D: Calculation of the integrals over $z$, $z^\prime$, and $q_z$ in the spin expression in Eq. (\ref{48})}
Let us consider the general form of the integrals which are to be evaluated:
\begin{eqnarray}\label{D1}
A = \int_{-\infty}^{\infty} e^{iQz} e^{-i a \sin kz} dz.
\end{eqnarray}
We now separate the interval of integration into parts. Each part has length $2\pi$. Then (\ref{D1}) transforms to
\begin{eqnarray}\label{D2}
A&=&\sum_{m = - \infty}^{\infty} \int_{-\pi/k}^{\pi/k}e^{iQ(z + 2\pi m/k)} e^{-i a \sin kz}dz \nonumber
\\
&=&\sum_{m = - \infty}^{\infty}e^{iQ 2\pi m/k } \int_{-\pi/k}^{\pi/k}e^{iQ z  } e^{-i a \sin kz}dz.
\end{eqnarray}
The first exponential in (\ref{D2}) has the representation \cite{LandauStat}
\begin{eqnarray}\label{D3}
\sum_{m = - \infty}^{  \infty}e^{iQ 2\pi m/k}  =  \sum_{p = - \infty}^{  \infty} \delta (\frac{Q}{k} - p)= k \sum_{p = - \infty}^{  \infty} \delta ( Q  - p k).
\end{eqnarray}
Therefore, for $A$, we obtain
\begin{eqnarray}\label{D4}
A = \sum_{p = - \infty}^{  \infty} \delta ( Q  - p k) J_p^* (a).
\end{eqnarray}
The application of (\ref{D4}) to the integrals will give the desired expression:
\begin{eqnarray}\label{D5}
\int dzdz^\prime dq_z e^{\Phi_l} = &(2\pi)^2&\sum_{u=-\infty}^{\infty}e^{iklt\left(1-\frac{p_{0z}+kl+ku}{\sqrt(p_{0\perp}^2+m^2+(p_{0z}+kl+ku)^2)}\right)}\nonumber
\\
&\times& J_{-u}(-2\alpha\beta)J_{u+l}(2\alpha\beta).
\end{eqnarray}
\section*{Appendix E: Calculation of the average in spin space}
In order to calculate spin four-vector we need to calculate the averages in the Dirac spinor space. Which are
\begin{eqnarray}\label{E1}
&&\frac{\bar u(p_{n})}{\sqrt{2\ep_{n}}}\gamma^5\gamma^\mu\frac{u(p_n)}{\sqrt{2\ep_n}}; 
\quad
\frac{\bar u(p_{n})}{\sqrt{2\ep_{n}}}\hat b \hat k \gamma^5\gamma^\mu\hat k \hat b\frac{u(p_n)}{\sqrt{2\ep_n}} ; \nonumber
\\
&&\frac{\bar u(p_{n})}{\sqrt{2\ep_{n}}}(\hat b \hat k \gamma^5 \gamma^\mu+\gamma^5\gamma^\mu\hat k \hat b)\frac{u(p_n)}{\sqrt{2\ep_n}}
\end{eqnarray}
Inserting the density matrix of the electron
\begin{equation}\label{E2}
\rho =u(p)\otimes\bar u(p)= \frac{1}{2}(\hat p + m)(1-\gamma^5\hat a),
\end{equation}
into Equation (\ref{E1}), we obtain
\begin{eqnarray}\label{E3}
\frac{\bar u(p_{n})}{\sqrt{2\ep_{n}}}\gamma^5\gamma^\mu\frac{u(p_n)}{\sqrt{2\ep_n}}&=&\frac{1}{\ep_n}\mathrm{Sp}\rho\gamma^5\gamma^\mu ;\nonumber
\\
\frac{\bar u(p_{n})}{\sqrt{2\ep_{n}}}\hat b \hat k \gamma^5\gamma^\mu\hat k \hat b\frac{u(p_n)}{\sqrt{2\ep_n}} &=&-\frac{k^\mu b^2}{\ep_n}\mathrm{Sp}\rho\gamma^5\hat k; 
\\
\frac{\bar u(p_{n})}{\sqrt{2\ep_{n}}}\gamma^5(\hat b \hat k \gamma^\mu+\gamma^\mu\hat k \hat b)\frac{u(p_n)}{\sqrt{2\ep_n}} &=& 
\frac{1}{\ep_n}\mathrm{Sp}\big(k^\mu\rho\gamma^5\hat b-b^\mu \rho\gamma^5 \hat k\big).\nonumber
\end{eqnarray}
Taking into account the fact that the trace of the product of the gamma matrices is nonzero only for even numbers of matrices, $\left(\gamma^{5}\right)^2=1$ and $\gamma^5$ anticommutes with $\gamma^\mu$ we have
\begin{eqnarray}\label{E4}
\frac{\bar u(p_{n})}{\sqrt{2\ep_{n}}}\gamma^5\gamma^\mu\frac{u(p_n)}{\sqrt{2\ep_n}}&=&\frac{m}{\ep_n}a^\mu;\nonumber
\\
\frac{\bar u(p_{n})}{\sqrt{2\ep_{n}}}\hat b \hat k \gamma^5\gamma^\mu\hat k \hat b\frac{u(p_n)}{\sqrt{2\ep_n}} &=&-\frac{mk^\mu b^2}{\ep_n}\fp{a}{k} ;
\\
\frac{\bar u(p_{n})}{\sqrt{2\ep_{n}}}\gamma^5(\hat b \hat k \gamma^\mu+\gamma^\mu\hat k \hat b)\frac{u(p_n)}{\sqrt{2\ep_n}} &=& \frac{m}{\ep_n}\big(k^\mu\fp{a}{b}-b^\mu \fp{a}{k}). \nonumber
\end{eqnarray}
\nocite{*}
\bibliography{quantum-1}

\end{document}